\documentclass[12pt,twoside,english,refpage,intoc,bibliography=totoc,index=totoc,BCOR7.5mm,captions=tableheading]{article}
\usepackage{lmodern}

\usepackage[T1]{fontenc}
\usepackage[latin9]{inputenc}
\usepackage{geometry}
\usepackage{color}
\usepackage{babel}
\usepackage{amsmath}
\usepackage{amssymb}
\usepackage[unicode=true,
 bookmarks=true,bookmarksnumbered=false,bookmarksopen=false,
 breaklinks=false,pdfborder={0 0 1},backref=false,colorlinks=true]
 {hyperref}

\makeatletter
\numberwithin{equation}{section}

\usepackage{babel}
\usepackage[figure]{hypcap}

\usepackage{fancyhdr}

\@ifundefined{extrarowheight}{\usepackage{array}}{}
\usepackage{braket}

\makeatother

\begin{document}
\title{More on stubs in open string field theory}
\author{Martin Schnabl and Georg Stettinger}
\maketitle
\begin{center}
{CEICO, Institute of Physics of the Czech Academy of Sciences, \\
 Na Slovance 2, 182 00 Prague 8, Czech Republic} 
\par\end{center}
\begin{abstract}
We continue our analysis of open string field theory based on $A_{\infty}$-algebras
obtained from Witten's theory by attaching stubs to the elementary
vertex. Classical solutions of the new theory can be obtained from
known analytic solutions in Witten's theory by applying a cohomomorphism.
In a previous work two such cohomomorphisms were found, one non-cyclic,
obtained from the homological perturbation lemma and another one by
geometric methods. Here we show that to first order in the stub length
the two resulting maps are related by a combination of a gauge transformation
and a term vanishing on-shell. We also extend our construction to
more general gauges and explicitly calculate the first few orders
of the new $A_{\infty}$-algebra solutions in the sliver frame. 
\end{abstract}
\newpage{}

\tableofcontents{}

\section{Introduction and motivation}

Open string field theory (OSFT) as introduced in \cite{Witten1986}
is described by the action 
\begin{equation}
S\left(\Psi\right)=\frac{1}{2}\omega\left(\Psi,Q\Psi\right)+\frac{1}{3}\omega\left(\Psi,m_{2}\left(\Psi,\Psi\right)\right)
\end{equation}
where $Q$ is the BRST-operator, $m_{2}$ is the Witten star product
and $\omega$ is the BPZ-product. Due to the simplicity of the action,
a lot of analytical methods have been developed \cite{Erler:2019vhl}
which enabled the discovery of classical analytical solutions, most
importantly the tachyon vacuum \cite{Schnabl:2005gv}. Despite those
successes, it is as well of interest to analyze modifications of the
theory. For once, we would like to have a formulation of OSFT which
structure is closer to that of closed string field theory (CSFT):
In this way we can not only gain more insight into the more complicated
CSFT, but also get a step closer to explicitly formulating a combined
open-closed SFT \cite{Zwiebach1998}. Moreover, we expect certain
singular behaviour of the Witten theory, for instance concerning identity
based solutions \cite{Arroyo2010,Erler:2019vhl}, to be ameliorated
in a modified version.

In \cite{Schnabl:2023dbv} a specific modification of OSFT is discussed
where stubs are attached to the Witten three-vertex\footnote{Stubs in open and closed string field theories have been introduced
by Zwiebach \cite{Zwiebach1993,Zwiebach1998} and further studied
in \cite{Moeller2011,Takezaki2019}. Toy models for stubs were considered
in \cite{Chiaffrino2022}. More recent works which deal with stubs
include \cite{Erbin2023,Erler2023}. }, i. e. 
\begin{equation}
m_{2}\left(\cdot,\cdot\right)\rightarrow M_{2}\left(\cdot,\cdot\right)=e^{-\lambda L_{0}}m_{2}\left(e^{-\lambda L_{0}}\cdot,e^{-\lambda L_{0}}\cdot\right).
\end{equation}
This new product is not associative, which makes it necessary to introduce
infinitely many higher products $M_{n}$ for $n\geq3$ to ensure gauge
invariance of the action. Those higher products are however only unique
up to a gauge choice. The whole set $Q,\,M_{2},\,M_{3},...$ then
forms a cyclic $A_{\infty}$-algebra. Explicitly, the higher products
were constructed using a slightly generalized version of homotopy
transfer resulting in the following definition: $M_{n}$ is equal
to the sum of all distinct, rooted, full planar binary trees with
$n$ leaves such that every leaf represents one input and the root
is the output. With every leaf there is one factor of $i=e^{-\lambda L_{0}}$
associated, with every node the product $m_{2}$, with every internal
line the homotopy $h$, which in Siegel gauge takes the form $h=\frac{e^{-2\lambda L_{0}}-1}{L_{0}}b_{0}$
and with the root $p=e^{-\lambda L_{0}}$. So for instance, 
\begin{align}
M_{3}\left(\Psi_{1,}\Psi_{2},\Psi_{3}\right)= & \,e^{-\lambda L_{0}}m_{2}\left(e^{-\lambda L_{0}}\Psi_{1},\frac{e^{-2\lambda L_{0}}-1}{L_{0}}b_{0}m_{2}\left(e^{-\lambda L_{0}}\Psi_{2},e^{-\lambda L_{0}}\Psi_{3}\right)\right)\nonumber \\
 & \,+\,e^{-\lambda L_{0}}m_{2}\left(\frac{e^{-2\lambda L_{0}}-1}{L_{0}}b_{0}m_{2}\left(e^{-\lambda L_{0}}\Psi_{1},e^{-\lambda L_{0}}\Psi_{2}\right),e^{-\lambda L_{0}}\Psi_{3}\right).
\end{align}
The higher vertices have a nice geometric interpretation: They consist
of all the string Feynman diagrams where the propagator is replaced
by an integral over strips of the form 
\begin{equation}
h=-\int_{0}^{2\lambda}dt\,e^{-tL_{0}}b_{0}=\frac{e^{-2\lambda L_{0}}-1}{L_{0}}b_{0}.
\end{equation}
Comparing with the standard Schwinger representation of the propagator
\begin{equation}
-\int_{0}^{\infty}dt\,e^{-tL_{0}}b_{0}=-\frac{b_{0}}{L_{0}}
\end{equation}
one sees that the vertices cover exactly those Riemann surfaces which
are missed by the ordinary Feynman diagrams after the inclusion of
stubs. This ensures that all Feynman diagrams including the new higher
elementary vertices generate a full single cover of the moduli space
of bordered Riemann surfaces.

Now the main interest in this stubbed theory lies in studying its
classical solutions and how to obtain them from solutions of the standard
Witten theory. Defining 
\begin{equation}
\mathbf{m}=\mathbf{Q}+\mathbf{m_{2}},\,\,\,\,\,\,\,\,\,\,\,\,\,\mathbf{M}=\mathbf{Q}+\mathbf{M_{2}}+\mathbf{M_{3}}+...
\end{equation}
as the coderivations encoding the DGA of Witten theory and the $A_{\infty}$-algebra
of the stubbed theory, respectively, the equations of motion can be
written concisely as 
\begin{equation}
\mathbf{m}\frac{1}{1-\Psi}=0,\,\,\,\,\,\,\,\,\,\,\,\,\,\,\,\,\mathbf{M}\frac{1}{1-\Psi'}=0.
\end{equation}
In \cite{Schnabl:2023dbv}, two cohomomorphisms $\mathbf{P}$ and
$\mathbf{F}$ were introduced and discussed in detail, both having
the property of mapping solutions of one theory to solutions of the
other. $\mathbf{P}$ is a non-linear map derived via the homological
perturbation lemma and explicitly given up to quadratic order by 
\begin{equation}
\pi_{1}\mathbf{P}\frac{1}{1-\Psi}=e^{-\lambda L_{0}}\Psi+e^{-\lambda L_{0}}m_{2}\left(\Psi,\frac{e^{-2\lambda L_{0}}-1}{L_{0}}b_{0}\Psi\right)+e^{-\lambda L_{0}}m_{2}\left(\frac{e^{-2\lambda L_{0}}-1}{L_{0}}b_{0}\Psi,e^{-2\lambda L_{0}}\Psi\right)+\mathcal{O}\left(\Psi^{3}\right).
\end{equation}

Now $\mathbf{P}$ obeys the chain map relation 
\begin{equation}
\mathbf{MP}=\mathbf{Pm},
\end{equation}
i. e. $\mathbf{P}$ intertwines between the two algebras. As a result,
it maps solutions of Witten theory to solutions of the stubbed theory,

\begin{equation}
\mathbf{M}\frac{1}{1-\pi_{1}\mathbf{P}\frac{1}{1-\Psi}}=0\,\,\,\text{if}\,\,\,\mathbf{m}\frac{1}{1-\Psi}=0,
\end{equation}
where the relation (\ref{eq:cohomo relation}) was used. In contrast,
$\mathbf{F}$ was derived by geometrical methods and reads 
\begin{align}
\pi_{1}\mathbf{F}\frac{1}{1-\Psi'}= & \,\Psi'+\int_{0}^{\lambda}dt\,\,(e^{-tL_{0}}m_{2}\left(e^{-tL_{0}}b_{0}\Psi',e^{-tL_{0}}\Psi'\right)+e^{-tL_{0}}m_{2}\left(e^{-tL_{0}}\Psi',e^{-tL_{0}}b_{0}\Psi'\right)\nonumber \\
 & \,-e^{-tL_{0}}b_{0}m_{2}\left(e^{-tL_{0}}\Psi',e^{-tL_{0}}\Psi'\right))+\mathcal{O}\left(\Psi'^{3}\right).
\end{align}
It fulfills the opposite intertwining relation 
\begin{equation}
\mathbf{FM}=\mathbf{mF}
\end{equation}
which gives rise to

\begin{equation}
\mathbf{m}\frac{1}{1-\pi_{1}\mathbf{F}\frac{1}{1-\Psi'}}=0\,\,\,\text{if\,\,\,}\mathbf{M}\frac{1}{1-\Psi'}=0
\end{equation}
hence it maps solutions of the stubbed theory to solutions of the
cubic theory. \footnote{The fact that they are naturally defined in the opposite direction
stems from the manifestly different derivation, see \cite{Schnabl:2023dbv}. }

A fundamental difference between the two occurs by examining the action.
Since the Witten action can be written as \cite{Vosmera:2020jyw}
\footnote{Here, $\Psi\left(t\right)$ is a smooth interpolation with the properties
$\Psi\left(0\right)=0$ and $\Psi\left(1\right)=\Psi.$} 
\begin{equation}
S\left(\Psi\right)=\int_{0}^{1}dt\,\omega\left(\pi_{1}\boldsymbol{\partial_{t}}\frac{1}{1-\Psi\left(t\right)},\pi_{1}\mathbf{m}\frac{1}{1-\Psi\left(t\right)}\right)\label{eq:Witten action}
\end{equation}
we would expect the stubbed action to read 
\begin{equation}
S'\left(\Psi\right)=\int_{0}^{1}dt\,\omega\left(\pi_{1}\boldsymbol{\partial_{t}}\frac{1}{1-\Psi\left(t\right)},\pi_{1}\mathbf{M}\frac{1}{1-\Psi\left(t\right)}\right)\label{eq:nice A inf action}
\end{equation}
and indeed, $S'$ is generated by the cohomomorphism $\mathbf{F}$
via the relation 
\begin{equation}
S'\left(\Psi\right)=S\left(\pi_{1}\mathbf{F}\frac{1}{1-\Psi}\right).
\end{equation}
On the contrary we get 
\begin{equation}
S\left(\pi_{1}\mathbf{P^{-1}}\frac{1}{1-\Psi}\right)=\int_{0}^{1}dt\,\omega\left(\pi_{1}\mathbf{P^{-1}}\boldsymbol{\partial_{t}}\frac{1}{1-\Psi\left(t\right)},\pi_{1}\mathbf{P^{-1}M}\frac{1}{1-\Psi\left(t\right)}\right)=:\tilde{S}\left(\Psi\right),\label{eq:strange action}
\end{equation}
where $\tilde{S}$ leads to the same equations of motion as $S'$
but takes a significantly different form. The reason is that $\mathbf{P}$
(and so also $\mathbf{P^{-1}}$) is not a cyclic cohomomorphism with
respect to $\omega$, hence it does not obey 
\begin{equation}
\omega\left(\mathbf{P}\cdot,\mathbf{P}\cdot\right)=\omega\left(\cdot,\cdot\right).
\end{equation}
We can also form the combined transformation $\mathbf{T}=:\mathbf{FP}$
which obeys 
\begin{equation}
\mathbf{Tm}=\mathbf{mT}
\end{equation}
and is therefore a symmetry of the Witten equations of motion. However,
it manifestly changes the action such that it cannot be a symmetry
of the full quantum theory, it rather behaves like some generalized
symmetry. Those rather surprising facts raise some interesting questions: 
\begin{enumerate}
\item How are the two actions physically related? And connected to that 
\item what is the physical meaning of the combined transformation $\mathbf{T}$? 
\item Is there a more general family of actions which leads to the expected
equations of motion? 
\end{enumerate}
Moreover, most of the known analytic solutions are formulated in the
sliver frame, so to study those explicitly in the stubbed theory we
have to ask: 
\begin{enumerate}
\item Can the whole construction of the higher products and the cohomomorphisms
be generalized to the sliver frame? 
\item How will explicit solutions of the stubbed theory look like? 
\item Can we infer some general structure of Maurer-Cartan elements of $A_{\infty}$-algebras? 
\end{enumerate}
Those questions shall be addressed and answered within this work.

\section{Systematic analysis of the intertwining cohomomorphism}

The purpose of this section is to gain a better understanding of the
cohomomorphisms $\mathbf{P}$ and $\mathbf{F}$. Therefore we find
a general strategy how to obtain intertwining cohomomorphisms and
then analyze $\mathbf{P}$ and $\mathbf{F}$ from this perspective.
We are looking for a non-linear field redefinition\footnote{Field redefinitions have been studied using cohomomorphisms already
in the context of open superstring field theory with the goal of relating
the Berkovits theory to the Munich construction, see \cite{Erler2015,Erler2015a,Erler2016}).
In our work we use them to find the field redefinition between the
stubbed theory and Witten theory.} of the form 
\begin{equation}
\Psi\rightarrow\Psi'=A_{1}\left(\Psi\right)+A_{2}\left(\Psi,\Psi\right)+A_{3}\left(\Psi,\Psi,\Psi\right)+...
\end{equation}
with $A_{n}$ being a collection of multi-linear maps which ensures
that $\Psi'$ is a classical solution of the stubbed theory as long
as $\Psi$ is a solution of Witten theory, i. e. 
\begin{equation}
\mathbf{M}\frac{1}{1-\Psi'}=0\,\,\,\,\,\,\,\text{if}\,\,\,\,\,\,\,\mathbf{m}\frac{1}{1-\Psi}=0.\label{eq:eoms}
\end{equation}
Since in the tensor algebra formalism finite transformations are encoded
in cohomomorphisms\footnote{For a short summary of the tensor algebra formalism see Appendix A.},
it is natural to package the $A_{n}$ into a cohomorphism $\mathbf{A}$
in the standard way s. t. 
\begin{equation}
\Psi'=\pi_{1}\mathbf{A}\frac{1}{1-\Psi}.
\end{equation}

Now applying (\ref{eq:cohomo relation}) 
\begin{equation}
\mathbf{M}\frac{1}{1-\Psi'}=\mathbf{M}\frac{1}{1-\pi_{1}\mathbf{A}\frac{1}{1-\Psi}}=\mathbf{MA}\frac{1}{1-\Psi}
\end{equation}
naively suggests to solve for $\mathbf{MA}=\mathbf{m}$, but that
is in general not possible since a coderivation composed with a cohomomorphism
does not yield a coderivation. However, the combination $\mathbf{A^{-1}MA}$
is always a coderivation and solving 
\begin{equation}
\mathbf{A^{-1}MA}=\mathbf{m}\,\,\,\,\,\,\,\text{or}\,\,\,\,\,\,\mathbf{MA}=\mathbf{Am}\label{eq:fundamental finite}
\end{equation}
still implies (\ref{eq:eoms}), hence this is the fundamental relation
we will try to solve. In the context of homotopy transfer it appeared
as the chain map relation and simply states that $\mathbf{A}$ acts
as an intertwiner between the two coderivations which define the algebras.
It is easy to see that the set of all intertwiners forms a vector
space.

\subsection{Infinitesimal treatment}

The object $\mathbf{M}\left(\lambda\right)$ is in fact a continuous
family of coderivations parametrized by the stub length $\lambda\in[0,\infty)$,
obeying $\mathbf{M}\left(0\right)=\mathbf{m}.$ Similarly, $\mathbf{A}\left(\lambda_{1},\lambda_{2}\right)$
is a continuous family with $\mathbf{A}\left(\lambda_{1},\lambda_{2}\right)=\mathbf{\mathbf{1}}$
for $\lambda_{1}=\lambda_{2}$, hence we can write (\ref{eq:fundamental finite})
more generally as 
\begin{equation}
\mathbf{M}\left(\lambda_{2}\right)\mathbf{A}\left(\lambda_{1},\lambda_{2}\right)=\mathbf{A}\left(\lambda_{1},\lambda_{2}\right)\mathbf{\mathbf{M}}\left(\lambda_{1}\right).\label{eq:more general fundamental finite}
\end{equation}

Suppose we want to know the infinitesimal cohomomorphism which takes
us from any fixed $\lambda$ to $\lambda+\delta\lambda:$ It will
take the form $\mathbf{A}\left(\lambda,\lambda+\delta\lambda\right)=\mathbf{\mathbf{1}}+\delta\lambda\mathbf{a}\left(\lambda\right)+\mathcal{O}\left(\delta\lambda^{2}\right)$
with $\mathbf{a}\left(\lambda\right)$ some coderivation. Plugging
into (\ref{eq:more general fundamental finite}) straightforwardly
yields 
\begin{equation}
\left[\mathbf{a}\left(\lambda\right),\mathbf{M}\left(\lambda\right)\right]=\frac{d}{d\lambda}\mathbf{M}\left(\lambda\right).\label{eq:fundamental inf}
\end{equation}
Since this is an equation of coderivations, it is sufficient to examine
the projection to one output. Choosing $\lambda=0$, (\ref{eq:fundamental inf})
acting on $n$ inputs becomes

\begin{align}
 & n=1:\,\,\,\,\,\,\,\,\,\,\,\,\,\,\,\,\,\,\pi_{1}\left[\mathbf{a_{1}\left(0\right),Q}\right]\pi_{1}=\mathbf{0}\\
 & n=2:\,\,\,\,\,\,\,\,\,\,\,\,\,\,\,\,\,\,\pi_{1}\mathbf{\left(\left[\mathbf{a_{1}}\left(0\right),m_{2}\right]+\left[a_{2}\left(0\right),\mathbf{Q}\right]\right)}\pi_{2}=\pi_{1}\left(\mathbf{-L_{0}m_{2}-m_{2}\mathbf{L_{0}}}\right)\pi_{2}\\
 & n=3:\,\,\,\,\,\,\,\,\,\,\,\,\,\,\,\,\,\,\pi_{1}\mathbf{\left(\left[\mathbf{a_{2}}\left(0\right),\mathbf{m_{2}}\right]+\left[a_{3}\left(0\right),\mathbf{Q}\right]\right)}\pi_{3}=-2\pi_{1}\mathbf{m_{2}\left(b_{0}\mathbf{m_{2}}\right)}\pi_{3}\\
 & n\geq4:\,\,\,\,\,\,\,\,\,\,\,\,\,\,\,\,\,\,\pi_{1}\left(\mathbf{\left[\mathbf{a_{n-1}}\left(0\right),\mathbf{m_{2}}\right]+\left[\mathbf{a_{n}}\left(0\right),\mathbf{Q}\right]}\right)\pi_{n}=\mathbf{0}
\end{align}

where $\mathbf{L_{0}}$ and $\mathbf{b_{0}}$ are the coderivations
associated to $L_{0}$ and $b_{0}$, see (\ref{eq:coderivation}).
At $\lambda=0$, $\frac{d}{d\lambda}M_{n}\left(\lambda\right)$ vanishes
for $n\geq4$ since it contains $n-2$ factors of $h$, which are
of order $\lambda$. It is straightforward to write down those equations
for any finite $\lambda.$

\subsection{Finite transformation }

To find the finite intertwiner $\mathbf{A}\left(\lambda\right)$,
we first need to solve (\ref{eq:fundamental inf}) for all $\lambda$,
which results in 
\begin{equation}
\left[\mathbf{a}_{1}\left(\lambda\right),\mathbf{M}_{n}\left(\lambda\right)\right]+\left[\mathbf{a}_{2}\left(\lambda\right),\mathbf{M}_{n-1}\left(\lambda\right)\right]+...+\left[\mathbf{a}_{n}\left(\lambda\right),\mathbf{Q}\right]=\frac{d}{d\lambda}\mathbf{M}_{n}\left(\lambda\right).\label{eq:fundamental all lambda}
\end{equation}
for all $n$. This equation has a structural resemblance of the parallel
transport equation in gauge theories \cite{Nakahara2003}. If we have
some matter field $\psi\left(x^{\mu}\right)$ defined along some curve
$\gamma$ in spacetime parametrized by $t$ and some gauge connection
$B_{\mu}$, then we say that $\psi$ is parallel transported along
$\gamma$ if it fulfills 
\begin{equation}
\frac{d\psi\left(\gamma\left(t\right)\right)}{dt}=-\frac{dx^{\mu}}{dt}B_{\mu}\psi\left(\gamma\left(t\right)\right).
\end{equation}
The solution is given by the path-ordered exponential 
\begin{equation}
\psi\left(\gamma\left(t\right)\right)=\mathcal{P}e^{-\int_{0}^{t}ds\,B_{\mu}\frac{dx^{\mu}}{ds}}\psi\left(\gamma\left(0\right)\right).
\end{equation}
In our case the ``connection'' $\mathbf{a}\left(\lambda\right)$
acts on $\mathbf{M}$ via a commutator. This implies that after integration
the path-ordered exponential has to act in the adjoint way, i. e.
\begin{equation}
\mathbf{M}\left(\lambda\right)=\mathcal{P}e^{\int_{0}^{\lambda}dt\,\mathbf{a}\left(t\right)}\mathbf{m}\mathcal{P}e^{-\int_{0}^{\lambda}dt\,\mathbf{a}\left(t\right)},
\end{equation}
hence we can identify the finite cohomomorphism as 
\begin{equation}
\mathbf{A}\left(\lambda\right)=\mathcal{P}e^{\int_{0}^{\lambda}dt\,\mathbf{a}\left(t\right)}.
\end{equation}
By expanding the exponential we can write the new solution\footnote{We thank Jakub Vošmera for useful discussions and suggestions on that.}
as 
\begin{align}
\pi_{1}\mathbf{A}\frac{1}{1-\Psi}= & \,\Psi+\int_{0}^{\lambda}dt\,\,a_{1}^{t}\Psi+\int_{0}^{\lambda}dt\,\,a_{2}^{t}\left(\Psi,\Psi\right)+\int_{0}^{\lambda}dt\,\,a_{3}^{t}\left(\Psi,\Psi,\Psi\right)+\int_{0}^{\lambda}dt\,\,a_{1}^{t}\left(\int_{0}^{t}ds\,\,a_{1}^{s}\Psi\right)\nonumber \\
 & \,+\int_{0}^{\lambda}dt\,\,a_{1}^{t}\left(\int_{0}^{t}ds\,\,a_{2}^{s}\left(\Psi,\Psi\right)\right)+\int_{0}^{\lambda}dt\,\,a_{2}^{t}\left(\left(\int_{0}^{t}ds\,\,a_{1}^{s}\Psi\right),\Psi\right)\nonumber \\
 & \,+\int_{0}^{\lambda}dt\,\,a_{2}^{t}\left(\Psi,\left(\int_{0}^{t}ds\,\,a_{1}^{s}\Psi\right)\right)+\int_{0}^{\lambda}dt\,\,a_{3}^{t}\left(\left(\int_{0}^{t}ds\,\,a_{1}^{s}\Psi\right),\Psi,\Psi\right)\nonumber \\
 & \,+\int_{0}^{\lambda}dt\,\,a_{3}^{t}\left(\Psi,\left(\int_{0}^{t}ds\,\,a_{1}^{s}\Psi\right),\Psi\right)+\int_{0}^{\lambda}dt\,\,a_{3}^{t}\left(\Psi,\Psi,\left(\int_{0}^{t}ds\,\,a_{1}^{s}\Psi\right)\right)\nonumber \\
 & \,+\int_{0}^{\lambda}dt\,\,a_{1}^{t}\left(\int_{0}^{t}ds\,\,a_{3}^{s}\left(\Psi,\Psi,\Psi\right)\right)+\int_{0}^{\lambda}dt\,\,a_{2}^{t}\left(\int_{0}^{t}ds\,\,a_{2}^{s}\left(\Psi,\Psi\right),\Psi\right)\nonumber \\
 & \,+\int_{0}^{\lambda}dt\,\,a_{2}^{t}\left(\Psi,\int_{0}^{t}ds\,\,a_{2}^{s}\left(\Psi,\Psi\right)\right)+\mathcal{O}\left(\Psi^{\otimes4}\right)+\mathcal{O}\left(\lambda^{3}\right).\label{eq:path-ordered exp}
\end{align}
To lighten the notation we have denoted the $t$-dependence of $a\left(t\right)$
by a corresponding superscript. In general it is non-trivial to solve
the integrals explicitly, the only cohomomorphism we have available
in integrated form is $\mathbf{P}$, which was constructed in a manifestly
finite way by the homological perturbation lemma. 

\subsection{The cyclic cohomomorphism $\mathbf{F}$}

It is instructive to check the linearized equations explicitly for
the two known cohomomorphisms $\mathbf{F}$ and $\mathbf{P}$. $\mathbf{F}$
was already derived in the infinitesimal form (see \cite{Schnabl:2023dbv})
which is given by 
\begin{equation}
f_{n}\left(\lambda\right)=\pi\left(\mathbf{-b_{0}M_{n}\left(\lambda\right)+M_{n}\left(\lambda\right)\mathbf{b_{0}}}\right)\pi_{n}\label{eq:inf F}
\end{equation}
for $n\geq2$ and $f_{1}=0$ (here $\mathbf{b_{0}}$ is the coderivation
associated to $b_{0}$). To be consistent with the previous analysis
we have to consider $\mathbf{F^{-1}}$ instead of $\mathbf{F}$ though,
but on the infinitesimal level this will result only in an overall
sign change. By expanding around $\lambda=0$ we see that only $f_{2}$
is non-vanishing and the relevant equations become\footnote{In the following $\left[\cdot,\cdot\right]$ will always denote a
commutator and $\left\{ \cdot,\cdot\right\} $ will always denote
an anticommutator, regardless of the Grassmannality of the entries. } 
\begin{equation}
\pi_{1}\left[\mathbf{Q},\mathbf{f_{2}}\left(0\right)\right]\pi_{2}=\,\pi_{1}\mathbf{\left(-\left[\mathbf{Q},b_{0}m_{2}\right]+\left[\mathbf{Q},m_{2}\mathbf{b_{0}}\right]\right)}\pi_{2}=\pi_{1}\mathbf{\left(-L_{0}m_{2}-m_{2}\mathbf{L_{0}}\right)}\pi_{2},\label{eq:F check 1}
\end{equation}
\begin{align}
\pi_{1}\left[\mathbf{m_{2}},\mathbf{f_{2}}\left(0\right)\right] & \pi_{3}=\,\pi_{1}\left(\mathbf{-m_{2}\left(b_{0}\mathbf{m_{2}}\right)+b_{0}m_{2}\mathbf{m_{2}}+m_{2}\left(\mathbf{m_{2}}\mathbf{b_{0}}\right)-m_{2}\mathbf{b_{0}m_{2}}}\right)\pi_{3}=-2\pi_{1}\mathbf{m_{2}\left(b_{0}\mathbf{m_{2}}\right)}\pi_{3}.\label{eq:F check 2}
\end{align}
The first equation follows in a simple way from $\left\{ Q,b_{0}\right\} =L_{0}$
and the Leibniz rule $\left\{ \mathbf{Q},\mathbf{m_{2}}\right\} =0$,
whereas the second one uses associativity $\left\{ \mathbf{m_{2}},\mathbf{m_{2}}\right\} =0$
and follows after expanding all the coderivations. \footnote{When two coderivations appear in parentheses it means that the first
one always has to act on the output of the second one, see Appendix.} The finite form of $\mathbf{F}$ is then given according to (\ref{eq:path-ordered exp})
by 
\begin{align}
\pi_{1}\mathbf{F}\frac{1}{1-\Psi'} & =\,\Psi'+\int_{0}^{\lambda}dt\,\,\Big(e^{-tL_{0}}m_{2}\left(e^{-tL_{0}}b_{0}\Psi',e^{-tL_{0}}\Psi'\right)+e^{-tL_{0}}m_{2}\left(e^{-tL_{0}}\Psi',e^{-tL_{0}}b_{0}\Psi'\right)\nonumber \\
 & \,\,\,\,\,\,-e^{-tL_{0}}b_{0}m_{2}\left(e^{-tL_{0}}\Psi',e^{-tL_{0}}\Psi'\right)\Big)+\mathcal{O}\left(\Psi'^{3}\right).
\end{align}

\subsection{The non-cyclic cohomomorphism $\mathbf{P}$}

In \cite{Schnabl:2023dbv}, $\mathbf{P}$ was derived from the homological
perturbation lemma as 
\begin{equation}
\mathbf{P}=\mathbf{p\left(1-m_{2}h\right)^{-1}}
\end{equation}
and given explicitly as a finite transformation which reads to the
first few orders 
\begin{align}
P_{1}\, & =p\nonumber \\
P_{2}\, & =pm_{2}\left(\cdot,h\cdot\right)+pm_{2}\left(h\cdot,ip\cdot\right)\nonumber \\
P_{3}\, & =pm_{2}\left(\cdot,hm_{2}\left(\cdot,h\cdot\right)\right)+pm_{2}\left(\cdot,hm_{2}\left(h\cdot,ip\cdot\right)\right)+pm_{2}\left(h\cdot,hm_{2}\left(ip\cdot,ip\cdot\right)\right)+pm_{2}\left(hm_{2}\left(\cdot,h\cdot\right),ip\cdot\right)\nonumber \\
 & \,\,\,\,\,\,+pm_{2}\left(hm_{2}\left(h\cdot,ip\cdot\right),ip\cdot\right)+pm_{2}\left(h\cdot,ipm_{2}\left(\cdot,h\cdot\right)\right)+pm_{2}\left(h\cdot,ipm_{2}\left(h\cdot,ip\cdot\right)\right)...\,\,\,\,\,\,.\label{eq:P explicit}
\end{align}
Here, the individual maps are given by 
\begin{equation}
i=p=e^{-\lambda L_{0}},\,\,\,\,\,\,\,\,\,\,h=\frac{e^{-2\lambda L_{0}}-1}{L_{0}}b_{0}.
\end{equation}
As it was argued and proven in \cite{Schnabl:2023dbv,Erbin2023},
$\mathbf{P}$ obeys the chain map relation $\mathbf{MP}=\mathbf{Pm}$
provided one assumes the side conditions $h^{2}=hi=ph=0$ as well
as $pi=1$. In practice this means that in the expansion of $\mathbf{P}$
every $pi$ appearing should be replaced by unity and every term containing
one of the side conditions should be set to zero. 

We are now interested in the infinitesimal form of $\mathbf{P}$,
i. e. taking the ``path-ordered logarithm''. Expanding around $\lambda=0$
as $\mathbf{P}=\mathbf{1}+\delta\lambda\mathbf{g}$ yields 
\begin{align}
g_{1}\, & =-L_{0}\nonumber \\
g_{2}\, & =-2\pi_{1}\mathbf{m_{2}\mathbf{b_{0}}}\pi_{2}\nonumber \\
g_{n\geq3}\, & =0
\end{align}
and the relevant equations become 
\begin{equation}
\pi_{1}\mathbf{\left[g_{1},Q\right]}\pi_{1}=-\pi_{1}\mathbf{\left[L_{0},Q\right]}\pi_{1}=0
\end{equation}
\begin{equation}
\pi_{1}\left(\mathbf{\left[\mathbf{g_{1}},m_{2}\right]+\left[g_{2},\mathbf{Q}\right]}\right)\pi_{2}=-\pi_{1}\left(\left[\mathbf{\mathbf{L_{0}},m_{2}}\right]+2\left[\mathbf{m_{2}\mathbf{b_{0}},\mathbf{Q}}\right]\right)\pi_{2}=\pi_{1}\left\{ \mathbf{\mathbf{L_{0}},m_{2}}\right\} \pi_{2}
\end{equation}
\begin{equation}
\pi_{1}\left(\left[\mathbf{g_{2}},\mathbf{m_{2}}\right]+\left[\mathbf{g_{3},}\mathbf{Q}\right]\right)\pi_{3}=-2\pi_{1}\left[\mathbf{\left(m_{2}b_{0}\right)},\mathbf{m_{2}}\right]\pi_{3}=-2\pi_{1}\mathbf{m_{2}}\left(\mathbf{b_{0}m_{2}}\right)\pi_{3}.
\end{equation}
Again, all of them can be checked straightforwardly by using the well-known
commutation relations of the operators that occur. However, we will
now use the structure of $\mathbf{f}$ and $\mathbf{g}$ to determine
a more general family of solutions of (\ref{eq:fundamental inf}).
It is worth pointing out that although $\mathbf{F}$ and $\mathbf{P}$
look quite similar when expanded around $\lambda=0$, their finite
versions are fundamentally different: While we have $\mathbf{P}$
available explicitly, $\mathbf{F}$ is only known as a path-ordered
exponential. In fact, both expressions are given as expansions in
the tensor algebra but to calculate the action of $\mathbf{F}$ we
need an extra expansion in the number of integrals. 

\subsection{The symmetry map $\mathbf{T}$\label{subsec:The-symmetry-map}}

At this point it is actually an interesting task to examine the combined
map $\mathbf{T=FP}$ in more detail. We know from $\mathbf{MP}=\mathbf{Pm}$
and $\mathbf{MF^{-1}}=\mathbf{F^{-1}m}$ that 
\begin{equation}
\mathbf{Tm=mT}
\end{equation}
holds, hence $\mathbf{T}$ commutes with the equations of motion of
the Witten theory. This is the generic condition for a symmetry of
the equations of motion, however, $\mathbf{T}$ does not preserve
the action, which manifestly changes (see (\ref{eq:strange action})).
Hence it cannot generate a gauge symmetry but rather some kind of
generalized symmetry that we will work out now. It is useful to linearize
$\mathbf{T}$ around $\lambda=0$: If $\mathbf{T}=\mathbf{1}+\delta\lambda\mathbf{t}$
then $\mathbf{t}$ is given by 
\begin{align}
t_{1}\, & =g_{1}-f_{1}=-L_{0}\nonumber \\
t_{2}\, & =g_{2}-f_{2}=-\pi_{1}\mathbf{\left\{ \mathbf{b_{0}},m_{2}\right\} }\pi_{2}\nonumber \\
t_{n\geq3}\, & =0\label{eq:infinitesimal FP}
\end{align}
and it induces the transformation 
\begin{equation}
\Psi\rightarrow\Psi-\delta\lambda\left(L_{0}\Psi+b_{0}m_{2}\left(\Psi,\Psi\right)+m_{2}\left(b_{0}\Psi,\Psi\right)+m_{2}\left(\Psi,b_{0}\Psi\right)\right).\label{eq:inf gauge + trivial trafo}
\end{equation}
Let us define a gauge parameter $\Lambda=b_{0}\Psi$, then (\ref{eq:inf gauge + trivial trafo})
can be rewritten as 
\begin{equation}
\Psi\rightarrow\Psi-\delta\lambda\left(Q\Lambda+m_{2}\left(\Lambda,\Psi\right)+m_{2}\left(\Psi,\Lambda\right)+b_{0}\left(Q\Psi+m_{2}\left(\Psi,\Psi\right)\right)\right).
\end{equation}
We see that the transformation we get is a combination of an infinitesimal
gauge transformation and a term being proportional to the equations
of motion. While the gauge transformation part was expected in a symmetry
of the equations of motion, the other part is more interesting and
more unconventional. It is this part which is responsible for the
change of the off-shell action, however it is also clear that the
value of the on-shell action is preserved. This was already conjectured
in \cite{Schnabl:2023dbv}, since this value has physical significance
and we did not expect to find a physically distinct solution by applying
$\mathbf{T}.$

Extending our analysis to higher orders in $\lambda$ quickly becomes
cumbersome and we leave this problem for the future. In principle
it is possible that the flow of $\mathbf{T}\left(\lambda\right)$
leaves the gauge orbit for a finite $\lambda$, although we do not
expect this for physical reasons, see \cite{Schnabl:2023dbv}.

\subsection{More general solution for the intertwiner\label{subsec:More-general-solution}}

Given $\mathbf{M}\left(\lambda\right)$, the most general solution
of (\ref{eq:fundamental finite}) is actually hard to describe explicitly.
However, we will now construct a more general family of intertwiners
that includes $\mathbf{F^{-1}}$, which was found already. The equations
(\ref{eq:fundamental all lambda}) are linear in $\mathbf{a}$ with
an inhomogeneity so once we found a solution we can add an arbitrary
solution of the homogenous equation 
\begin{equation}
\left[\mathbf{a}_{1}\left(\lambda\right),\mathbf{M}_{n}\left(\lambda\right)\right]+\left[\mathbf{a}_{2}\left(\lambda\right),\mathbf{M}_{n-1}\left(\lambda\right)\right]+...+\left[\mathbf{a}_{n}\left(\lambda\right),\mathbf{Q}\right]=0.\label{eq:homogenous eq}
\end{equation}
This is the infinitesimal form of 
\begin{equation}
\mathbf{AM}=\mathbf{MA},
\end{equation}
the equation for the symmetry discussed in section \ref{subsec:The-symmetry-map}.
For $n=1$, it reduces to 
\begin{equation}
\left[a_{1},Q\right]=0
\end{equation}
which is solved by various operators. The simplest example are arbitrary
linear combinations of arbitrary products of Virasoro operators. Another
example would be $a_{1}\left(\cdot\right)=m_{2}\left(\Phi,\cdot\right)$
where $\Phi$ is any $Q$-closed string field of ghost number zero.
For simplicity we will focus just on linear combinations of Virasoros.
Motivated by the form of (\ref{eq:infinitesimal FP}) we make the
ansatz 
\begin{equation}
a_{1}=\sum_{k}v_{k}L_{k},\,\,\,\,\,\,\,\,\,a_{n\geq2}=\sum_{k}v_{k}\pi_{1}\mathbf{\left\{ \mathbf{b_{k}},M_{n}\right\} }\pi_{n}.
\end{equation}
This can be directly inserted into (\ref{eq:homogenous eq}) acting
on $n$ elements: 
\begin{align}
 & \sum_{k}v_{k}\pi_{1}\left(\left[\mathbf{\mathbf{L_{k}},M_{n}}\right]+\sum_{i=2}^{n-1}\mathbf{\left[\left(b_{k}\mathbf{M_{i}}\right),\mathbf{M_{n+1-i}}\right]+\left[\mathbf{\left(M_{i}b_{k}\right)},\mathbf{M_{n+1-i}}\right]+\left[\left(b_{k}M_{n}\right),\mathbf{Q}\right]+\left[\mathbf{\left(M_{n}b_{k}\right)},\mathbf{Q}\right]}\right)\pi_{n}\nonumber \\
=\, & \sum_{k}v_{k}\pi_{1}\left(\mathbf{L_{k}M_{n}-M_{n}\mathbf{L_{k}}}+\sum_{i=2}^{n-1}\mathbf{b_{k}\mathbf{M_{i}}\mathbf{M_{n+1-i}}-\mathbf{M_{n+1-i}}\left(b_{k}\mathbf{M_{i}}\right)+\mathbf{M_{i}b_{k}}\mathbf{M_{n+1-i}}}\right)\pi_{n}\nonumber \\
 & -\sum_{k}v_{k}\pi_{1}\left(\mathbf{M_{n+1-i}}\left(\mathbf{M_{i}b_{k}}\right)+\mathbf{b_{k}M_{n}\mathbf{Q}-Qb_{k}M_{n}+M_{n}\mathbf{b_{k}}\mathbf{Q}-QM_{n}\mathbf{b_{k}}}\right)\pi_{n}\nonumber \\
=\, & \sum_{k}v_{k}\pi_{1}\left(\mathbf{b_{k}QM_{n}-M_{n}\mathbf{Qb_{k}}}+\sum_{i=2}^{n-1}\mathbf{b_{k}\mathbf{M_{i}}\mathbf{M_{n+1-i}}-\mathbf{M_{n+1-i}}\left(b_{k}\mathbf{M_{i}}\right)}\right)\pi_{n}\nonumber \\
 & +\sum_{k}v_{k}\pi_{1}\left(\mathbf{M_{i}b_{k}}\mathbf{M_{n+1-i}}-\mathbf{M_{n+1-i}}\left(\mathbf{M_{i}b_{k}}\right)+\mathbf{b_{k}M_{n}\mathbf{Q}-QM_{n}\mathbf{b_{k}}}\right)\pi_{n}.
\end{align}
Using the $A_{\infty}$-relation

\begin{equation}
\left\{ \mathbf{M_{n}},\mathbf{Q}\right\} +\sum_{i=2}^{n-1}\mathbf{M_{i}}\mathbf{M_{n+1-i}}=0
\end{equation}
this expression can be shown to vanish, which proves our ansatz to
be correct.

We can also show that our solution for $\mathbf{a}$ is again a combination
of a gauge transformation and field redefinition proportional to the
equations of motion: If we define analogously to section \ref{subsec:The-symmetry-map}
\begin{equation}
\Lambda=\sum_{k}v_{k}b_{k}\Psi
\end{equation}
then we have 
\begin{align}
\pi_{1}\mathbf{a}\frac{1}{1-\Psi}\, & =Q\Lambda+\sum_{n=2}^{\infty}M_{n}\left(\Lambda,\Psi^{\otimes n-1}\right)+M_{n}\left(\Psi,\Lambda,\Psi^{\otimes n-2}\right)+...+M_{n}\left(\Psi^{\otimes n-1},\Lambda\right)\nonumber \\
 & +\sum_{k}v_{k}b_{k}\left(Q\Psi+\sum_{n=2}^{\infty}M_{n}\left(\Psi^{\otimes n}\right)\right).
\end{align}

To sum up, the family 
\begin{equation}
a_{1}=\sum_{k}v_{k}L_{k},\,\,\,\,\,\,\,\,\,a_{n\geq2}=\sum_{k}\pi_{1}\left(\left(v_{k}+\delta_{0k}\right)\mathbf{b_{k}M_{n}}+\left(v_{k}-\delta_{0k}\right)\mathbf{M_{n}\mathbf{b_{k}}}\right)\pi_{n}\label{eq:more general intertw}
\end{equation}
provides an infinitesimal intertwiner for all possible $v_{k}$ where
$v_{k}=0$ corresponds to $\mathbf{F^{-1}}$$.$

\subsection{Cyclicity and invariance of the action}

In this section we shall analyze under which conditions the transformation
$\mathbf{A}$ generates the expected $A_{\infty}$-action $S'$ (\ref{eq:nice A inf action}).
For that, $\mathbf{A}$ not only needs to be an intertwiner but also
be compatible with the symplectic form $\omega$, i. e. the last equation
in 
\begin{align}
S\left(\Psi\right)\, & =\int_{0}^{1}dt\,\omega\left(\pi_{1}\boldsymbol{\partial_{t}}\frac{1}{1-\Psi\left(t\right)},\pi_{1}\mathbf{m}\frac{1}{1-\Psi\left(t\right)}\right)\nonumber \\
 & =\int_{0}^{1}dt\,\omega\left(\pi_{1}\boldsymbol{\partial_{t}}\left(\mathbf{A^{-1}A}\frac{1}{1-\Psi\left(t\right)}\right),\pi_{1}\mathbf{A^{-1}MA}\frac{1}{1-\Psi\left(t\right)}\right)\nonumber \\
 & =\int_{0}^{1}dt\,\omega\left(\pi_{1}\boldsymbol{\partial_{t}}\mathbf{A^{-1}}\frac{1}{1-\pi_{1}\mathbf{A}\frac{1}{1-\Psi\left(t\right)}},\pi_{1}\mathbf{A^{-1}M}\frac{1}{1-\pi_{1}\mathbf{A}\frac{1}{1-\Psi\left(t\right)}}\right)\nonumber \\
 & =\int_{0}^{1}dt\,\omega\left(\pi_{1}\mathbf{A^{-1}}\boldsymbol{\partial_{t}}\frac{1}{1-\Psi'\left(t\right)},\pi_{1}\mathbf{A^{-1}M}\frac{1}{1-\Psi'\left(t\right)}\right)\nonumber \\
 & =\int_{0}^{1}dt\,\omega\left(\pi_{1}\boldsymbol{\partial_{t}}\frac{1}{1-\Psi'\left(t\right)},\pi_{1}\mathbf{M}\frac{1}{1-\Psi'\left(t\right)}\right)=S'\left(\Psi\right)
\end{align}
needs to be true. Note that $\partial_{t}$ and $\mathbf{A}$ commute
since $\mathbf{A}$ does not depend on $t$. 

Cyclicity of a cohomomorphism is actually a delicate question: In
\cite{Vosmera:2020jyw} it is stated as the condition 
\begin{equation}
\omega\left(\mathbf{A}\cdot,\mathbf{A}\cdot\right)=\omega\left(\cdot,\cdot\right)
\end{equation}
but one has to be precise on what type of elements it is supposed
to act. It is quite clear that the relation is too restrictive to
act on arbitrary elements: One would get 
\begin{align}
 & \,\omega\left(A_{1}\left(\Psi_{1}\right)+A_{2}\left(\Psi_{1},\Psi_{2}\right)+A_{3}\left(\Psi_{1},\Psi_{2},\Psi_{3}\right)+\cdots,A_{1}\left(\phi_{1}\right)+A_{2}\left(\phi_{1},\phi_{2}\right)+A_{3}\left(\phi_{1},\phi_{2},\phi_{3}\right)+\cdots\right)\nonumber \\
= & \,\omega\left(\Psi_{1},\phi_{1}\right)
\end{align}
which would imply (at least in the case where $A_{1}$ is invertible,
which is equivalent to $\mathbf{A}$ being invertible) that every
output of $A_{n\geq2}$ is orthogonal to any possible $\Psi.$ Since
$\omega$ is non-degenerate, we would conclude that the $A_{n\geq2}$
all have to be identically zero, which is not what we want. Even if
we only allow group-like inputs, i. e. elements of the form $\frac{1}{1-\Psi}$,
the same argument shows that the two inputs have to be identical.
By looking at (\ref{eq:Witten action}) however we see that we need
a generalization of that by allowing coderivations to act on the group-like
inputs. The equation 
\begin{align}
 & \omega\Big(\pi_{1}\left(\mathbf{A_{1}d_{1}}\left(\Psi\right)+\mathbf{A_{2}}\mathbf{d_{1}}\left(\Psi^{\otimes2}\right)+\mathbf{A_{1}d_{2}}\left(\Psi^{\otimes2}\right)+\cdots\right),\nonumber \\
 & \,\,\,\,\,\,\,\,\,\pi_{1}\left(\mathbf{A_{1}d'_{1}}\left(\Psi\right)+\mathbf{A_{2}}\mathbf{d'_{1}}\left(\Psi^{\otimes2}\right)+\mathbf{A_{1}d'_{2}}\left(\Psi^{\otimes2}\right)+\cdots\right)\Big)\nonumber \\
= & \,\omega\left(\pi_{1}\mathbf{d}\frac{1}{1-\Psi},\pi_{1}\mathbf{d}'\frac{1}{1-\Psi}\right)
\end{align}
actually makes sense also for non-trivial $A_{n\geq2}$ because we
can have 
\begin{equation}
\omega\left(A_{1}\pi_{1}\mathbf{d}\frac{1}{1-\Psi},A_{1}\pi_{1}\mathbf{d}'\frac{1}{1-\Psi}\right)=\omega\left(\pi_{1}\mathbf{d}\frac{1}{1-\Psi},\pi_{1}\mathbf{d}'\frac{1}{1-\Psi}\right)
\end{equation}
while the higher terms of a given order in $\Psi$ cancel each other,
even if $\mathbf{d}\neq\mathbf{d'}.$ For an infinitesimal $\mathbf{A}=\mathbf{1}+\epsilon\mathbf{a}+\mathcal{O}\left(\epsilon^{2}\right)$
it boils down to the condition that $\mathbf{a}$ is cyclic coderivation.
To sum up, we define a cohomomorphism $\mathbf{A}$ to be cyclic with
respect to $\omega$ if 
\begin{equation}
\omega\left(\pi_{1}\mathbf{Ad}\frac{1}{1-\Psi},\pi_{1}\mathbf{Ad}'\frac{1}{1-\Psi}\right)=\omega\left(\pi_{1}\mathbf{d}\frac{1}{1-\Psi},\pi_{1}\mathbf{d}'\frac{1}{1-\Psi}\right)
\end{equation}
for arbitrary $\mathbf{d}$, $\mathbf{d'}$ and $\Psi$.

With this definition we can immediately analyze the cyclicity properties
of $\mathbf{F^{-1}}$ and $\mathbf{P}$: With the infinitesimal $\mathbf{f}$
given by (\ref{eq:inf F}) we get 
\begin{align}
\omega\left(\Psi_{1},f_{n}\left(\Psi_{2},...,\Psi_{n+1}\right)\right)\, & =-\omega\left(\Psi_{1},b_{0}M_{n}\left(\Psi_{2},...,\Psi_{n+1}\right)\right)+\omega\left(\Psi_{1},M_{n}\left(\mathbf{b_{0}}\left(\Psi_{2},...,\Psi_{n+1}\right)\right)\right)\nonumber \\
 & =-\omega\left(b_{0}\Psi_{1},M_{n}\left(\Psi_{2},...,\Psi_{n+1}\right)\right)-\omega\left(M_{n}\left(\Psi_{1},\mathbf{b_{0}}\left(\Psi_{2},...,\Psi_{n}\right)\right),\Psi_{n+1}\right)\nonumber \\
 & \,\,\,\,\,\,\,\,\,\,-\omega\left(M_{n}\left(\Psi_{1},...,\Psi_{n}\right),b_{0}\Psi_{n+1}\right)\nonumber \\
 & =-\omega\left(M_{n}\left(b_{0}\Psi_{1},\Psi_{2},...,\Psi_{n}\right),\Psi_{n+1}\right)-\omega\left(M_{n}\left(\Psi_{1},\mathbf{b_{0}}\left(\Psi_{2},...,\Psi_{n}\right)\right),\Psi_{n+1}\right)\nonumber \\
 & \,\,\,\,\,\,\,\,\,\,\,+\omega\left(b_{0}M_{n}\left(\Psi_{1},...,\Psi_{n}\right),\Psi_{n+1}\right)\nonumber \\
 & =-\omega\left(M_{n}\mathbf{b_{0}}\left(\Psi_{1},\Psi_{2},...,\Psi_{n}\right),\Psi_{n+1}\right)+\omega\left(b_{0}M_{n}\left(\Psi_{1},...,\Psi_{n}\right),\Psi_{n+1}\right)\nonumber \\
 & =-\omega\left(f_{n}\left(\Psi_{1},...,\Psi_{n}\right)\Psi_{n+1}\right),\label{eq:cyclicity proof}
\end{align}
hence $\mathbf{f}$ is cyclic and so is the finite version 
\begin{equation}
\mathbf{F^{-1}}=\mathcal{P}e^{\int_{0}^{\lambda}dt\,\mathbf{f}\left(t\right)}.
\end{equation}
In contrast, for $\mathbf{P}$ we already see at first order that
$f_{1}'=-L_{0}$ is not cyclic because the sign does not match: 
\begin{equation}
-\omega\left(\Psi_{1},L_{0}\Psi_{2}\right)=-\omega\left(L_{0}\Psi_{1},\Psi_{2}\right).
\end{equation}
This implies that $P_{1}=e^{-\lambda L_{0}}$ is not ``unitary''
with respect to the BPZ-product as it would be required for a cyclic
cohomomorphism.

From the results of section 2.6 we also deduce that $\mathbf{F^{-1}}$
is not unique as a cyclic intertwiner: Taking a BPZ-odd choice for
$a_{1}$ in (\ref{eq:more general intertw}), i. e. demanding $v_{k}=-\left(-1\right)^{k}v_{-k}$
leads to a cyclic coderivation and in turn to a cyclic cohomomorphism
by a similar argument as in (\ref{eq:cyclicity proof}).

\section{Generalized stubs}

We now want to go a step further and allow for more general stub operators,
especially non-BPZ-even ones. The motivation behind that is that we
want to apply our construction to explicit analytic solutions of OSFT.
While it is in principle straightforward to do that, we face a technical
problem: The most important solutions, like for instance the tachyon
vacuum (\cite{Schnabl:2005gv,Erler:2009uj}), are formulated in the
sliver frame in terms of the $KBc$-algebra. The action of $i$, $p$
and $h$ would take us outside the $KBc$-algebra and is therefore
impractical for actual calculations. It would be much more natural
to use the sliver frame analogue of the stub operator, i. e. replace
$e^{-\lambda L_{0}}$ by $e^{-\lambda\mathcal{\mathcal{L}}_{0}}$.
We will first discuss general aspects of non-BPZ-even stub operators
and provide a careful treatment of the operator $e^{-\lambda\mathcal{\mathcal{L}}_{0}}$
in section \ref{subsec:The-sliver-frame}.

\subsection{Algebraic aspects}

Let us consider a generalized stub operator of the form 
\begin{equation}
e^{-\lambda\sum v_{k}L_{k}}=:e^{-\lambda L}
\end{equation}
with some real coefficients $v_{k}$. An important example is given
by the family 
\begin{equation}
L=L_{t}=L_{0}+2\sum_{k=1}^{\infty}\frac{\left(-1\right)^{k+1}}{4k^{2}-1}e^{-2tk}L_{2k}\label{eq:L interpolation}
\end{equation}
which interpolates between the Siegel gauge and sliver gauge stub:
For $t=0$ we get $\mathcal{\mathcal{L}}_{0}$, whereas in the limit
of $t\rightarrow\infty$ we recover $L_{0}$. The most important new
algebraic aspect is that $e^{-\lambda L}$ is not BPZ-even since in
general $L^{*}\neq L$. Hence, the naive choice $p=i=e^{-\lambda L}$
would not result in cyclic products, we need to define 
\begin{equation}
p=e^{-\lambda L^{*}}
\end{equation}
instead. While this small change seems innocuous at first sight, it
also affects the Hodge-Kodaira relation and therefore our possible
choices of $h$, which we use to construct the higher vertices. 

To motivate our general construction of $h$ let us first consider
the special case of $L=\mathcal{\mathcal{L}}_{0}$, postponing the
discussion of potential geometrical subtleties to section \ref{subsec:The-sliver-frame}.
The right-hand side of 
\begin{equation}
hQ+Qh=ip-1\label{eq:Hodge Kodaira}
\end{equation}
evaluates to 
\begin{equation}
e^{-\lambda\mathcal{\mathcal{L}}_{0}}e^{-\lambda\mathcal{\mathcal{L}}_{0}^{*}}-1=e^{\left(e^{-\lambda}-1\right)\left(\mathcal{\mathcal{L}}_{0}+\mathcal{L}_{0}^{*}\right)}-1\label{eq:h umformung}
\end{equation}
using the algebraic relations \cite{Schnabl:2005gv} 
\begin{equation}
x^{\mathcal{L}_{0}}y^{\mathcal{L}_{0}^{*}}=\left(\frac{1}{1+\frac{x}{y}-x}\right)^{\mathcal{L}_{0}^{*}}\left(\frac{1}{1+\frac{y}{x}-y}\right)^{\mathcal{L}_{0}},\,\,\,\,\,\,\,\,\,\,\,x^{\mathcal{L}_{0}^{*}}x^{\mathcal{L}_{0}}=e^{\left(1-\frac{1}{x}\right)\left(\mathcal{\mathcal{L}}_{0}+\mathcal{L}_{0}^{*}\right)}.\label{eq:algebraic relations}
\end{equation}
A natural choice for $h$ would now be 
\begin{equation}
h_{\hat{\mathcal{B}}_{0}}=\frac{e^{\left(e^{-\lambda}-1\right)\left(\mathcal{\mathcal{L}}_{0}+\mathcal{L}_{0}^{*}\right)}-1}{\left(\mathcal{\mathcal{L}}_{0}+\mathcal{L}_{0}^{*}\right)}\left(\mathcal{\mathcal{B}}_{0}+\mathcal{B}_{0}^{*}\right)=-\int_{0}^{1-e^{-\lambda}}dt\,\left(\mathcal{\mathcal{B}}_{0}+\mathcal{B}_{0}^{*}\right)e^{-t\left(\mathcal{\mathcal{L}}_{0}+\mathcal{L}_{0}^{*}\right)}.\label{eq:h in B_hat gauge}
\end{equation}
This expression is manifestly non-singular: If $\mathcal{\mathcal{L}}_{0}+\mathcal{L}_{0}^{*}$
yields zero on some state (which would for example formally the case
for the sliver state) then no pole is produced. It corresponds to
the propagator in $\hat{\mathcal{B}}_{0}$-gauge 
\begin{equation}
-\int_{0}^{\infty}dt\,\left(\mathcal{\mathcal{B}}_{0}+\mathcal{B}_{0}^{*}\right)e^{-t\left(\mathcal{\mathcal{L}}_{0}+\mathcal{L}_{0}^{*}\right)}=-\frac{\mathcal{\mathcal{B}}_{0}+\mathcal{B}_{0}^{*}}{\mathcal{\mathcal{L}}_{0}+\mathcal{L}_{0}^{*}}\equiv-\frac{\hat{\mathcal{B}}_{0}}{\hat{\mathcal{L}}_{0}}.
\end{equation}
From an algebraic perspective, this propagator is quite convenient
and leads to a simple set of vertices.

The solution for $h_{\hat{\mathcal{B}}_{0}}$ relied heavily on the
special algebraic properties of $\mathcal{\mathcal{L}}_{0}$ and $\mathcal{L}_{0}^{*}$.
To find a solution for a generic $L$, most importantly $L_{t}$ defined
above, we can take 
\begin{equation}
h_{\hat{B}}=-\int_{0}^{\lambda}dt\,e^{-tL}\left(B+B^{*}\right)e^{-tL^{*}}\label{eq:h in B_hat gauge-1}
\end{equation}
with $B$ defined as 
\begin{equation}
B=\underset{k}{\sum}v_{k}B_{k}.
\end{equation}
For $L=\mathcal{\mathcal{L}}_{0}$ and $B=\mathcal{\mathcal{B}}_{0}$
it reduces to (\ref{eq:h in B_hat gauge}). Plugging into the Hodge-Kodaira
relation and using $\left\{ Q,B\right\} =L$ we get 
\begin{equation}
Qh_{\hat{B}}+h_{\hat{B}}Q=-\int_{0}^{\lambda}dt\,e^{-tL}\left(L+L^{*}\right)e^{-tL^{*}}=\int_{0}^{\lambda}dt\,\frac{d}{dt}\left(e^{-tL}e^{-tL^{*}}\right)=e^{-\lambda L}e^{-\lambda L^{*}}-1=ip-1
\end{equation}
as desired. Hence we succeeded to find a well-defined solution for
the homotopy for any generalized stub $e^{-\lambda L}$.

\subsection{Geometric aspects}

As we discussed in \cite{Schnabl:2023dbv} already, we have to ensure
that the Feynman diagrams constructed out of the vertices and the
propagator provide a full single cover of the moduli space of bordered
punctured Riemann surfaces. This implies that the higher elementary
vertices must include precisely those surfaces which are missed after
attaching the stubs. What changes compared to \cite{Schnabl:2023dbv}
is that $e^{-\lambda L}$ is not the time evolution operator in radial
quantization anymore and induces a non-trivial distortion of the worldsheet
surface \cite{Kiermaier:2007jg}.

We have seen above that every choice of stub operator $e^{-\lambda L}$
is naturally associated to a gauge condition given by 
\begin{equation}
B\Psi=0.
\end{equation}
However, the homotopy $h_{\hat{B}}$ we constructed in (\ref{eq:h in B_hat gauge-1})
is part of the propagator in $\hat{B}$-gauge, i. e. where the gauge
condition 
\begin{equation}
\left(B+B^{*}\right)\Psi=0
\end{equation}
is imposed. We could ask now if there is also a choice for $h$ that
corresponds to $B$-gauge directly. After all, analytic solutions
have been found in $\mathcal{\mathcal{B}}_{0}$-gauge, not $\hat{\mathcal{B}}_{0}$-gauge,
where they become singular. Hence an $h$ that corresponds to sliver
gauge (i. e. $\mathcal{\mathcal{B}}_{0}\Psi=0$) would seem more natural.
To answer that, we will use some geometric input from computing amplitudes
in general linear $B$-gauges.

In \cite{Kiermaier:2007jg} a condition on $B=\underset{k}{\sum}v_{k}B_{k}$
was given that ensures that all tree level amplitudes can be computed
unambiguously. In terms of the vector field $v\left(\xi\right)=\underset{k}{\sum}v_{k}\xi^{k+1}$
this condition reads 
\begin{equation}
\text{Re}\left(\overline{\xi}v\left(\xi\right)\right)>0\,\,\,\,\,\,\,\,\,\text{for}\,\,\,\,\,\,\,\,\,\mid\xi\mid=1.\label{eq:regularity condition}
\end{equation}
It is for example obeyed for the family 
\begin{equation}
B_{t}=:e^{tL_{0}}\mathcal{B}_{0}e^{-tL_{0}}=b_{0}+2\sum_{k=1}^{\infty}\frac{\left(-1\right)^{k+1}}{4k^{2}-1}e^{-2tk}b_{2k}
\end{equation}
associated to the interpolation (\ref{eq:L interpolation}) as long
as $t$ is strictly greater than zero. This means that for sliver
gauge the condition is marginally violated, see section \ref{subsec:The-sliver-frame}.
For this section we will assume that (\ref{eq:regularity condition})
holds for our choice of $v_{k}$. The propagator in a general $B$-gauge
was derived in \cite{Kiermaier:2007jg} to be 
\begin{equation}
-\frac{B^{*}}{L^{*}}Q\frac{B}{L}\,\,\,\,\text{on odd ghost number states,\,\,\,\,\,}-\frac{B}{L}Q\frac{B^{*}}{L^{*}}\,\,\,\,\text{on even ghost number states},\label{eq:B_t propagator}
\end{equation}
so it contains two Schwinger parameters instead of one. The result
is an infinite overcounting of the moduli space, every surface is
now additionally integrated over from zero to infinity. The reason
why the theory is still unitary and produces the right values for
amplitudes is the presence of $Q$ in the propagator, which cancels
the overcounting.

To derive the correct form of $h_{B}$ in $B$-gauge let us analyze
the on-shell four-amplitude in Witten theory as well as in the stubbed
theory. In Witten theory there is no elementary 4-vertex and the whole
amplitude is given by the Feynman region: 
\begin{equation}
\mathcal{A}_{4}=-\omega\left(\Psi_{1},m_{2}\left(\Psi_{2},\frac{B}{L}Q\frac{B^{*}}{L^{*}}m_{2}\left(\Psi_{3},\Psi_{4}\right)\right)\right)+\text{perm.}\label{eq:Witten amplitude}
\end{equation}
Here, perm. stands for the t-channel contribution obtained by a cyclic
permutation where $\frac{B}{L}Q\frac{B^{*}}{L^{*}}m_{2}$ acts on
$\Psi_{2}$ and $\Psi_{3}$. In the stubbed theory we have to sum
the Feynman region and the vertex region: 
\begin{equation}
\mathcal{A}_{4}=-\omega\left(\Psi_{1},M_{2}\left(\Psi_{2},\frac{B}{L}Q\frac{B^{*}}{L^{*}}M_{2}\left(\Psi_{3},\Psi_{4}\right)\right)\right)+\text{perm.}+\omega\left(\Psi_{1},M_{3}\left(\Psi_{2},\Psi_{3},\Psi_{4}\right)\right),
\end{equation}
where 
\begin{equation}
M_{3}\left(\cdot,\cdot,\cdot\right)=\,e^{-\lambda L^{*}}m_{2}\left(e^{-\lambda L}\cdot,h_{B}m_{2}\left(e^{-\lambda L}\cdot,e^{-\lambda L}\cdot\right)\right)\,+\,e^{-\lambda L^{*}}m_{2}\left(h_{B}m_{2}\left(e^{-\lambda L}\cdot,e^{-\lambda L}\cdot\right),e^{-\lambda L}\cdot\right).
\end{equation}
If we only focus on the s-channel (including the part of $M_{3}$
that ``extends'' to the s-channel) the expression becomes 
\begin{align}
\mathcal{A}_{4s}=-\, & \omega\left(\Psi_{1},e^{-\lambda L^{*}}m_{2}\left(e^{-\lambda L}\Psi_{2},e^{-\lambda L}\frac{B}{L}Q\frac{B^{*}}{L^{*}}e^{-\lambda L^{*}}m_{2}\left(e^{-\lambda L}\Psi_{3},e^{-\lambda L}\Psi_{4}\right)\right)\right)\nonumber \\
 & +\omega\left(e^{-\lambda L}\Psi_{1},m_{2}\left(e^{-\lambda L}\Psi_{2},h_{B}m_{2}\left(e^{-\lambda L}\Psi_{3},e^{-\lambda L}\Psi_{4}\right)\right)\right)
\end{align}
Since we take the external states to be on-shell, the stub operators
acting directly on $\Psi_{i}$ do not matter and we get 
\begin{align}
 & \mathcal{A}_{4s}=-\omega\left(\Psi_{1},m_{2}\left(\Psi_{2},e^{-\lambda L}\frac{B}{L}Q\frac{B^{*}}{L^{*}}e^{-\lambda L^{*}}m_{2}\left(\Psi_{3},\Psi_{4}\right)\right)\right)+\omega\left(\Psi_{1},m_{2}\left(\Psi_{2},h_{B}m_{2}\left(\Psi_{3},\Psi_{4}\right)\right)\right).
\end{align}
Comparing with (\ref{eq:Witten amplitude}) would give us $h_{B}$
when acting on a ghost number two state as 
\begin{align}
 & e^{-\lambda L}\frac{B}{L}Q\frac{B^{*}}{L^{*}}e^{-\lambda L^{*}}-\frac{B}{L}Q\frac{B^{*}}{L^{*}}
\end{align}
up to $Q$-exact terms. The sliver gauge propagator was dependent
on the ghost number of the input; if we demand that this property
should also hold for the homotopy we arrive at 
\begin{equation}
h_{B}=\left(e^{-\lambda L}\frac{B}{L}Q\frac{B^{*}}{L^{*}}e^{-\lambda L^{*}}-\frac{B}{L}Q\frac{B^{*}}{L^{*}}\right)P_{+}+\left(e^{-\lambda L}\frac{B^{*}}{L^{*}}Q\frac{B}{L}e^{-\lambda L^{*}}-\frac{B^{*}}{L^{*}}Q\frac{B}{L}\right)P_{-},\label{eq:h in Schnabl gauge}
\end{equation}
where $P_{+}$ ($P_{-}$) is the projector on states of even (odd)
ghost number. Now we can verify that $h_{B}$ also obeys the Hodge-Kodaira
relation (\ref{eq:Hodge Kodaira}). Note that the dependence on the
ghost number is crucial for that to work. We point out that this construction
of $h_{B}$ was purely governed by the consistency of the on-shell
amplitudes. We see that it is natural for $h_{B}$ to be in the same
gauge as the propagator which is used to compute amplitudes. An interesting
point is that (\ref{eq:h in Schnabl gauge}) can contain poles if
$L$ or $L^{*}$ give zero on some state, in contrast to $h_{\hat{\mathcal{B}}_{0}}$
(\ref{eq:h in B_hat gauge}). This is particularly important in the
sliver frame limit and will be discussed in section \ref{subsec:The-sliver-frame}.

To sum up, we are now able to construct the higher products in exactly
the same way as in \cite{Schnabl:2023dbv} but with $i=e^{-\lambda L}$,
$p=e^{-\lambda L^{*}}$ and $h$ equal to (\ref{eq:h in Schnabl gauge})
or (\ref{eq:h in B_hat gauge-1}), depending on the purposes.

\subsection{Intertwining cohomomorphisms}

Let us see now what changes regarding the intertwiners if we are working
with generalized stubs. Actually the construction of $\mathbf{P}$
is very simple: One can just use the modified homotopy transfer formula
of \cite{Schnabl:2023dbv} and replace $i$, $p$ and $h$ by their
generalized counterparts defined above. The only necessary algebraic
ingredients were the chain map relations $Qi=iQ$, $Qp=pQ$ as well
as (\ref{eq:Hodge Kodaira}), which we have shown to be true. Hence
we conclude that $\mathbf{P}$ constructed this way obeys the perturbed
chain map relation 
\begin{equation}
\mathbf{Pm}=\mathbf{MP}
\end{equation}
as desired.

To find a cyclic intertwiner that also preserves the form of the action
is a bit more involved: Motivated by the explicit form of $\mathbf{f\left(\lambda\right)}$
in the standard frame we propose the following ansatz: 
\begin{equation}
f_{1}=0\,\,\,\,\,\,\,\,\,\,\,\,\,\,\text{and}\,\,\,\,\,\,\,\,\,\,\,\,\,f_{n}\left(\lambda\right)=\pi_{1}\left(\mathbf{X^{*}M_{n}\left(\lambda\right)-M_{n}\left(\lambda\right)}\mathbf{X}\right)\pi_{n}\label{eq:cyclic intertwiner}
\end{equation}
Cyclicity then follows automatically by (\ref{eq:cyclicity proof})
with $X$ replacing $b_{0}$. Looking at equations (\ref{eq:F check 1}),
(\ref{eq:F check 2}) in combination with (\ref{eq:fundamental inf})
we can deduce the following conditions on $X$: 
\begin{align}
\left\{ Q,X\right\}  & \,=L,\\
X+X^{*} & \,=-\frac{d}{d\lambda}h\mid_{\lambda=0}
\end{align}
In $\hat{B}$-gauge it is actually simple because 
\begin{equation}
-\frac{d}{d\lambda}h_{\hat{B}}\mid_{\lambda=0}=\hat{B}
\end{equation}
and the natural choice $X=$$B$ provides a solution.

For $B$- gauge the situation is slightly more complicated, we get
\begin{equation}
-\frac{d}{d\lambda}h_{B}\mid_{\lambda=0}=\left(BQ\frac{B^{*}}{L^{*}}+\frac{B}{L}QB^{*}\right)P_{+}+\left(L\frac{B^{*}}{L^{*}}Q\frac{B}{L}+\frac{B^{*}}{L^{*}}Q\frac{B}{L}L^{*}\right)P_{-}.
\end{equation}
To solve both conditions we can define 
\begin{equation}
X=BQ\frac{B^{*}}{L^{*}}P_{+}+L\frac{B^{*}}{L^{*}}Q\frac{B}{L}P_{-}.\label{eq:X}
\end{equation}
To prove that our ansatz is indeed correct we need to insert into
(\ref{eq:fundamental all lambda}): 
\begin{equation}
\pi_{1}\left(\mathbf{\left[X^{*}M_{n},\mathbf{Q}\right]-\left[M_{n}\mathbf{X},\mathbf{Q}\right]}+\sum_{i=2}^{n-1}\left[\left(\mathbf{X^{*}}\mathbf{M_{n+1-i}}\right),\mathbf{M_{i}}\right]-\left[\mathbf{M_{n+1-i}X},\mathbf{M_{i}}\right]\right)\pi_{n}=\frac{d}{d\lambda}M_{n}
\end{equation}
The r. h. s. consists of two parts: One where the derivative acts
on the stubs and one where it acts on the homotopy. The action on
the stubs just brings down a factor of $-L$ or $-L^{*}$ whereas
the second part consists of all tree diagrams with one internal line
replaced by $\frac{d}{d\lambda}h_{B}\left(\lambda\right)$. The replaced
line divides the tree into two subtrees with $i$ and $n+1-i$ leaves,
respectively. The sum of those subtrees form the products $M_{i}$
and $M_{n+1-i}$ again such that we can write 
\begin{equation}
\frac{d}{d\lambda}M_{n}=\pi_{1}\left(\mathbf{-L^{*}M_{n}-M_{n}\boldsymbol{L}}-\sum_{i=2}^{n-1}\mathbf{M_{n+1-i}},\left(\left(\mathbf{X+X^{*}}\right)\mathbf{M_{i}}\right)\right)\pi_{n}.
\end{equation}
Indeed, the leaf of $M_{n+1-i}$ and the root of $M_{i}$ combine
to give 
\begin{equation}
e^{-\lambda L}\left(-X-X^{*}\right)e^{-\lambda L^{*}}=e^{-\lambda L}\frac{d}{d\lambda}h_{B}\mid_{\lambda=0}e^{-\lambda L^{*}}=\frac{d}{d\lambda}h_{B}\left(\lambda\right)
\end{equation}
as desired. We can now manipulate the l. h. s. using the $A_{\infty}$-relation
\begin{equation}
\left\{ \mathbf{M_{n}},\mathbf{Q}\right\} +\sum_{i=2}^{n-1}\mathbf{M_{n+1-i}}\mathbf{M_{i}}=0
\end{equation}
and get 
\begin{align}
 & \pi_{1}\Big(\mathbf{X^{*}\left\{ M_{n},\mathbf{Q}\right\} -\left\{ Q,X^{*}\right\} M_{n}-M_{n}\left\{ \mathbf{X},\mathbf{Q}\right\} +\left\{ M_{n},\mathbf{Q}\right\} \mathbf{X}}\nonumber \\
+\,\, & \sum_{i=2}^{n-1}\mathbf{X^{*}\mathbf{M_{n+1-i}}\mathbf{M_{i}}-\mathbf{M_{i}}\left(X^{*}\mathbf{M_{n+1-i}}\right)}-\mathbf{M_{n+1-i}X}\mathbf{M_{i}}+\mathbf{M_{i}\left(M_{n+1-i}X\right)\Big)}\pi_{n}\nonumber \\
=\,\, & \pi_{1}\left(\mathbf{-L^{*}M_{n}-M_{n}L}-\sum_{i=2}^{n-1}\mathbf{M_{n+1-i}},\left(\left(\mathbf{X+X^{*}}\right)\mathbf{M_{i}}\right)\right)\pi_{n}
\end{align}
and hence exactly the r. h. s. The finite cohomomorphism can now be
computed again as the path-ordered exponential:\footnote{Observe that $\mathbf{F}$ is now defined for simplicity in the same
direction as $\mathbf{P}$, so it corresponds to $\mathbf{F^{-1}}$
in \cite{Schnabl:2023dbv}.} 

\begin{equation}
\mathbf{F}=\mathcal{P}e^{\int_{0}^{\lambda}dt\,\mathbf{f}\left(t\right)}.
\end{equation}

\subsection{The sliver frame limit\label{subsec:The-sliver-frame}}

As already stated above, the sliver gauge condition marginally violates
the regularity condition (\ref{eq:regularity condition}) which has
the following geometric reason: The stub operator $e^{-\lambda L_{t}},$
which is also used in the Schwinger representation of the propagator,
does not attach just a rectangular strip, but a more general surface
to the world sheet. For $t=0$ this distortion of the strip becomes
singular in the sense that the string midpoint is pushed to an infinite
distance. This means that every stub $e^{-\lambda\mathcal{L}_{0}}$
independently of $\lambda$ covers an infinitely long region on the
Riemann surface. The same problem also concerns the operator $e^{-\lambda\left(\mathcal{\mathcal{L}}_{0}+\mathcal{L}_{0}^{*}\right)}$.
This raises the question of potential singularities and it also makes
it less obvious which region of moduli space will be covered by the
higher vertices. Moreover, there is also a caveat related to the Schwinger
representation: Since naively one would expect 
\begin{equation}
\frac{1}{L_{t}}=\underset{\Lambda\rightarrow\infty}{\text{lim}}\int_{0}^{\Lambda}dt\,e^{-tL_{t}}=\underset{\Lambda\rightarrow\infty}{\text{lim}}\frac{1}{L_{t}}\left(1-e^{-\Lambda L_{t}}\right),
\end{equation}
we must ensure that the second term gives zero contribution, i. e.
it must produce a surface on the boundary of the moduli space. It
has been shown that this is the case for $t$ strictly bigger than
zero but not for $t=0$ \cite{Kiermaier:2007jg}.

There are indeed singularities showing up in the sliver frame limit:
If we look at the homotopy $h_{\mathcal{B}_{0}}$(\ref{eq:h in Schnabl gauge})
it takes the form 
\begin{equation}
h_{\mathcal{B}_{0}}=\left(e^{-\lambda\mathcal{L}_{0}}\frac{\mathcal{B}_{0}}{\mathcal{L}_{0}}Q\frac{\mathcal{B}_{0}^{*}}{\mathcal{L}_{0}^{*}}e^{-\lambda\mathcal{L}_{0}^{*}}-\frac{\mathcal{B}_{0}}{\mathcal{L}_{0}}Q\frac{\mathcal{B}_{0}^{*}}{\mathcal{L}_{0}^{*}}\right)P_{+}+\left(e^{-\lambda\mathcal{L}_{0}}\frac{\mathcal{B}_{0}^{*}}{\mathcal{L}_{0}^{*}}Q\frac{\mathcal{B}_{0}}{\mathcal{L}_{0}}e^{-\lambda\mathcal{L}_{0}^{*}}-\frac{\mathcal{B}_{0}^{*}}{\mathcal{L}_{0}^{*}}Q\frac{\mathcal{B}_{0}}{\mathcal{L}_{0}}\right)P_{-}.
\end{equation}
Within the $KBc$-algebra there exist a couple of states on which
this expression is singular: While at ghost number zero there are
no singularities showing up, at ghost number one we have $e^{-\frac{K}{2}}cKe^{-\frac{K}{2}}$,
$e^{-\frac{K}{2}}Kce^{-\frac{K}{2}}$, $e^{-\frac{K}{2}}cKBce^{-\frac{K}{2}}$
and $e^{-\frac{K}{2}}ce^{-\frac{K}{2}}$ and at ghost number two $e^{-\frac{K}{2}}cKcKe^{-\frac{K}{2}}$
and $e^{-\frac{K}{2}}KcKce^{-\frac{K}{2}}$ (and linear combinations
thereof). The appearance of the zero-momentum tachyon in this list
might seem surprising since it has $\mathcal{L}_{0}$-eigenvalue minus
one, but the operator $e^{-\lambda\mathcal{L}_{0}^{*}}$ creates a
level zero state out of it, see section \ref{subsec:Action-of P}.
One should notice that for the ghost number two states, it is not
$\frac{1}{\mathcal{L}_{0}^{*}}$ but $\frac{1}{\mathcal{L}_{0}}$
on the left side of the expression which creates the singularity.
We can increase the range of definition of $h_{\mathcal{B}_{0}}$
to all $Q$-closed states, which includes $e^{-\frac{K}{2}}cKBce^{-\frac{K}{2}}$,
$e^{-\frac{K}{2}}cKcKe^{-\frac{K}{2}}$ and $e^{-\frac{K}{2}}KcKce^{-\frac{K}{2}}$
by the following trick: We rewrite the sliver gauge propagator as
\begin{equation}
\frac{\mathcal{B}_{0}^{*}}{\mathcal{L}_{0}^{*}}Q\frac{\mathcal{B}_{0}}{\mathcal{L}_{0}}P_{-}\cong\left(1-Q\frac{\mathcal{B}_{0}^{*}}{\mathcal{L}_{0}^{*}}\right)\frac{\hat{\mathcal{B}}_{0}}{\hat{\mathcal{L}}_{0}}\left(1-\frac{\mathcal{B}_{0}}{\mathcal{L}_{0}}Q\right)P_{-}\label{eq:propagator rewriting}
\end{equation}
and likewise for even ghost numbers. This expression is manifestly
BPZ-even and produces well-defined results on all $Q$-closed states
while being equivalent to the original version on all non-problematic
states\footnote{A nice side effect is that in this way we can construct an interpolation
between $h_{\mathcal{B}_{0}}$ and $h_{\hat{\mathcal{B}}_{0}}$: If
we define $h_{int}\left(\alpha\right)=\left(1-\alpha Q\frac{\mathcal{B}_{0}^{*}}{\mathcal{L}_{0}^{*}}\right)\frac{\hat{\mathcal{B}}_{0}}{\hat{\mathcal{L}}_{0}}\left(1-\alpha\frac{\mathcal{B}_{0}}{\mathcal{L}_{0}}Q\right)P_{-}+\left(1-\alpha Q\frac{\mathcal{B}_{0}}{\mathcal{L}_{0}}\right)\frac{\hat{\mathcal{B}}_{0}}{\hat{\mathcal{L}}_{0}}\left(1-\alpha\frac{\mathcal{B}_{0}^{*}}{\mathcal{L}_{0}^{*}}Q\right)P_{+}$
then $h_{int}\left(0\right)=h_{\hat{\mathcal{B}}_{0}}$ and $h_{int}\left(1\right)=h_{\mathcal{B}_{0}}$.}. Now the only true and non-curable singularities essentially occur
at the states $e^{-\frac{K}{2}}cKe^{-\frac{K}{2}}$, $e^{-\frac{K}{2}}Kce^{-\frac{K}{2}}$
and $e^{-\frac{K}{2}}ce^{-\frac{K}{2}}.$

They are however not a consequence of adding stubs, they stem from
the fact that the sliver gauge propagator itself has poles on those
states. The difference is that now a part of these pole contributions
is moved to the internal vertices and hence also creates singularities
in the equations of motion. Our strategy for making sense of the stubbed
theory in the sliver frame and for computing analytic solutions will
be to use the interpolating stub $e^{-\lambda L_{t}}$ and understand
it in the limit $t\rightarrow0$. This means for the geometric interpretation,
$\mathcal{L}_{0}$ $\left(\mathcal{B}_{0}\right)$ should be replaced
by $L_{t}$ $\left(B_{t}\right)$ in every expression while in the
end we let $t$ go to zero. From an algebraic point of view, this
is unproblematic for the $h_{\mathcal{\hat{B}}_{0}}$-vertices, where
no poles appear. The $h_{\mathcal{B}_{0}}$-vertices that were motivated
from the calculations of amplitudes are also fine as long as they
are restricted to on-shell states because typical representatives
of the cohomology do not include the above-mentioned problematic $KBc$-states.
It would be interesting if the range of definition of $h_{\mathcal{B}_{0}}$
can be extended to the full Hilbert space by including suitable projectors
and treat the problematic states separately. We will leave this problem
for future work and in this paper just analyze the singularities that
appear case by case. Moreover, so far it has not been proven in general
that all amplitudes can be defined consistently in sliver gauge and
a full proof of this statement lies beyond the scope of this paper
as well, some useful references include \cite{Kiermaier2008,Rastelli2008}. 

To proceed, we now want to give an argument why the inclusion of higher
vertices is necessary from a geometrical viewpoint, directly in the
sliver frame. Let us once again consider the s-channel contribution
to the on-shell four-point amplitude: 
\begin{equation}
\mathcal{A}_{4s}=-\omega\left(\Psi_{1},m_{2}\left(\Psi_{2},e^{-\lambda\mathcal{L}_{0}}\frac{\mathcal{B}_{0}}{\mathcal{L}_{0}}Q\frac{\mathcal{B}_{0}^{*}}{\mathcal{L}_{0}^{*}}e^{-\lambda\mathcal{L}_{0}^{*}}m_{2}\left(\Psi_{3},\Psi_{4}\right)\right)\right).
\end{equation}
Using Schwinger parameters we can write that as 
\begin{equation}
\mathcal{A}_{4s}=\int_{0}^{\infty}dt\int_{0}^{\infty}ds\,\omega\left(m_{2}\left(\Psi_{1},\Psi_{2}\right),e^{-\left(\lambda+t\right)\mathcal{L}_{0}}\mathcal{B}_{0}Q\mathcal{B}_{0}^{*}e^{-\left(\lambda+s\right)\mathcal{L}_{0}^{*}}m_{2}\left(\Psi_{3},\Psi_{4}\right)\right).
\end{equation}
Now we use the fact that $Q$ annihilates the on-shell states $\Psi_{3}$
and $\Psi_{4}$ and get 
\begin{equation}
\mathcal{A}_{4s}=\int_{0}^{\infty}dt\int_{0}^{\infty}ds\,\omega\left(m_{2}\left(\Psi_{1},\Psi_{2}\right),e^{-\left(\lambda+t\right)\mathcal{L}_{0}}\mathcal{B}_{0}\mathcal{L}_{0}^{*}e^{-\left(\lambda+s\right)\mathcal{L}_{0}^{*}}m_{2}\left(\Psi_{3},\Psi_{4}\right)\right).
\end{equation}
We see that the $\mathcal{L}_{0}^{*}$-insertion can be written as
a derivative which localizes the $s$-integral:
\begin{equation}
\mathcal{A}_{4s}=\underset{\Lambda\rightarrow\infty}{\text{lim}}\int_{0}^{\infty}dt\,\omega\left(m_{2}\left(\Psi_{1},\Psi_{2}\right),e^{-\left(\lambda+t\right)\mathcal{L}_{0}}\mathcal{B}_{0}\left(e^{-\lambda\mathcal{L}_{0}^{*}}-e^{-\left(\lambda+\Lambda\right)\mathcal{L}_{0}^{*}}\right)m_{2}\left(\Psi_{3},\Psi_{4}\right)\right).
\end{equation}
As mentioned already, the cut-off term containing $e^{-\Lambda\mathcal{L}_{0}^{*}}$
is not guaranteed to yield a vanishing contribution but we will first
of all focus on the first term. The next step is to use the state-operator
correspondence and write the star product in the form outlined in
\cite{Schnabl:2005gv,Schnabl2003}: 
\begin{equation}
\Psi_{1}\left(0\right)\Ket{0}*\Psi_{2}\left(0\right)\Ket{0}=\left(\frac{8}{9}\right)^{h_{1}+h_{2}}e^{\text{ln}\frac{2}{3}\mathcal{L}_{0}^{*}}\Psi_{1}\left(\frac{1}{\sqrt{3}}\right)\Psi_{2}\left(-\frac{1}{\sqrt{3}}\right)\Ket{0},
\end{equation}
where the $h_{i}$ are the conformal weights of the primary fields
$\Psi_{i}$. Since we take our external states to be on-shell, we
can omit the prefactor and get for the first term 
\begin{equation}
\mathcal{A}_{4s}^{\left(1\right)}=\int_{0}^{\infty}dt\,\Bra{0}\Psi_{2}\left(\sqrt{3}\right)\Psi_{1}\left(-\sqrt{3}\right)e^{-\left(\lambda+t-\text{ln}\frac{2}{3}\right)\mathcal{L}_{0}}\mathcal{B}_{0}e^{-\left(\lambda-\text{ln}\frac{2}{3}\right)\mathcal{L}_{0}^{*}}\Psi_{3}\left(\frac{1}{\sqrt{3}}\right)\Psi_{4}\left(-\frac{1}{\sqrt{3}}\right)\Ket{0}.
\end{equation}
To commute the two exponentials we can use the formula (\ref{eq:algebraic relations})
and arrive at 
\begin{equation}
\mathcal{A}_{4s}^{\left(1\right)}=\int_{0}^{\infty}dt\,\Bra{0}\Psi_{2}\left(\sqrt{3}\right)\Psi_{1}\left(-\sqrt{3}\right)\mathcal{B}_{0}\left(\frac{1}{1+e^{-t}-\frac{2}{3}e^{-\lambda-t}}\right)^{\mathcal{L}_{0}^{*}}\left(\frac{1}{1+e^{t}-\frac{2}{3}e^{-\lambda}}\right)^{\mathcal{L}_{0}}\Psi_{3}\left(\frac{1}{\sqrt{3}}\right)\Psi_{4}\left(-\frac{1}{\sqrt{3}}\right)\Ket{0}.
\end{equation}
The operator $x^{\mathcal{L}_{0}}$ is the scaling operator in the
sliver frame and acts on the upper-half-plane coordinates as $z\rightarrow\text{tan}\left(x\,\text{arctan}z\right)$.
Similarly, the $\mathcal{L}_{0}^{*}$-exponential can be made acting
to the left where it transforms the coordinates as $z\rightarrow\text{cot}\left(x\,\text{arctan}\frac{1}{z}\right)$.
The scaling of the operators can be omitted again and we get 
\begin{align}
\mathcal{A}_{4s}^{\left(1\right)}=\int_{0}^{\infty}dt\,\, & \Bra{0}\Psi_{2}\left(\text{cot}\left(\frac{\frac{\pi}{6}}{1+e^{-t}-\frac{2}{3}e^{-\lambda-t}}\right)\right)\Psi_{1}\left(\text{cot}\left(-\frac{\frac{\pi}{6}}{1+e^{-t}-\frac{2}{3}e^{-\lambda-t}}\right)\right)\nonumber \\
 & \mathcal{B}_{0}\Psi_{3}\left(\text{tan}\left(\frac{\frac{\pi}{6}}{1+e^{t}-\frac{2}{3}e^{-\lambda}}\right)\right)\Psi_{4}\left(\text{tan}\left(-\frac{\frac{\pi}{6}}{1+e^{t}-\frac{2}{3}e^{-\lambda}}\right)\right)\Ket{0}.
\end{align}
This is our final expression for the $s$-channel contribution in
terms of a four-point function dependent on one real modulus $t$.
This four-point function is some function of the cross-ratio of the
insertion points given by 
\begin{equation}
c_{s}=\frac{\left(z_{1}-z_{2}\right)\left(z_{3}-z_{4}\right)}{\left(z_{1}-z_{3}\right)\left(z_{2}-z_{4}\right)}=4\frac{\text{cot}\left(\frac{\frac{\pi}{6}}{1+e^{-t}-\frac{2}{3}e^{-\lambda-t}}\right)\text{tan}\left(\frac{\frac{\pi}{6}}{1+e^{t}-\frac{2}{3}e^{-\lambda}}\right)}{\left(\text{cot}\left(\frac{\frac{\pi}{6}}{1+e^{-t}-\frac{2}{3}e^{-\lambda-t}}\right)+\text{tan}\left(\frac{\frac{\pi}{6}}{1+e^{t}-\frac{2}{3}e^{-\lambda}}\right)\right)^{2}}.\label{eq:cross ratio}
\end{equation}
$c_{s}$ is a useful parameter of the moduli space of four-punctured
disks so by analyzing its range we can see which portion of the moduli
space is covered \cite{Ohmori2018}. 

Let us first consider Witten theory with $\lambda=0$: $c_{s}\left(t\right)$
is now a monotonically decreasing function with $c_{s}\left(0\right)=\frac{1}{2}$
and $\underset{t\rightarrow\infty}{\text{lim}}c_{s}\left(t\right)=0$.
This is an expected result: We consider just one specific ordering
of the operators here and choosing the standard locations $0$, $1$
and $\infty$ for three of them, $c_{s}$ is just given by the second
location $z_{2}$ and should therefore lie between $0$ and $1$ .
Hence the portion of the moduli space we expect to be covered is the
unit interval and the s-channel covers half of it. The t-channel contribution
can be simply found by a cyclic permutation $z_{i}\rightarrow z_{i+1}$
modulo $4$ and we get 
\begin{equation}
c_{t}=\frac{\left(z_{2}-z_{3}\right)\left(z_{4}-z_{1}\right)}{\left(z_{2}-z_{4}\right)\left(z_{3}-z_{1}\right)}=1-c_{s}
\end{equation}
so indeed the other half of the unit interval is covered. 

Now let us see what happens if we add stubs: For $\lambda>0$, $c_{s}\left(t\right)$
is still a monotonically decreasing function with $\underset{t\rightarrow\infty}{\text{lim}}c_{s}\left(t\right)=0$
but $c_{s}\left(0\right)<\frac{1}{2}$. This means that the interval
$\left(c_{s}\left(0\right),1-c_{s}\left(0\right)\right)$ is not covered
by the Feynman diagrams and adding higher vertices is necessary also
from a geometrical point of view. We can see that more explictly by
setting $t$ to zero in (\ref{eq:cross ratio}) to get 
\begin{equation}
c_{s}\left(t=0\right)=\text{sin}^{2}\left(\frac{e^{\lambda}}{1-3e^{\lambda}}\frac{\pi}{2}\right)
\end{equation}
This function is monotically decreasing, which means the uncovered
region gets bigger as the stub length is increased. An interesting
point is that in the limit of infinitely long stubs we get $c_{s}\left(t=0\right)=\frac{1}{4}$,
hence the Feynman region covers precisely half of the moduli space. 

Finally we want to analyze the cut-off term of the Schwinger parametrization
given by 
\begin{equation}
\mathcal{A}_{4s}^{\left(2\right)}=-\underset{\Lambda\rightarrow\infty}{\text{lim}}\int_{0}^{\infty}dt\,\omega\left(m_{2}\left(\Psi_{1},\Psi_{2}\right),e^{-\left(\lambda+t\right)\mathcal{L}_{0}}\mathcal{B}_{0}e^{-\left(\lambda+\Lambda\right)\mathcal{L}_{0}^{*}}m_{2}\left(\Psi_{3},\Psi_{4}\right)\right).
\end{equation}
Going through the same steps as before we can calculate the cross-ratio
and get 
\begin{equation}
c_{s}=4\frac{\text{cot}\left(\frac{\frac{\pi}{6}}{1+e^{-t+\Lambda}-\frac{2}{3}e^{-\lambda-t}}\right)\text{tan}\left(\frac{\frac{\pi}{6}}{1+e^{t-\Lambda}-\frac{2}{3}e^{-\lambda-\Lambda}}\right)}{\left(\text{cot}\left(\frac{\frac{\pi}{6}}{1+e^{-t+\Lambda}-\frac{2}{3}e^{-\lambda-t}}\right)+\text{tan}\left(\frac{\frac{\pi}{6}}{1+e^{t-\Lambda}-\frac{2}{3}e^{-\lambda-\Lambda}}\right)\right)^{2}}.
\end{equation}
In the limit of $\Lambda\rightarrow\infty$ we have $c_{s}\rightarrow0$,
hence the cut-off term indeed yields a contribution only at the boundary
of the moduli space, as expected. 

\section{The tachyon vacuum in the stubbed theory}

In this section we want to apply the cohomomorphisms we found on the
most important classical solution of Witten theory, namely the tachyon
vacuum. It is explicitly given by \cite{Schnabl:2005gv,Erler:2019vhl}
\begin{equation}
\Psi_{TV}=e^{-\frac{K}{2}}c\frac{KB}{1-e^{-K}}ce^{-\frac{K}{2}}
\end{equation}
with the elements of the $KBc$-algebra defined for instance in \cite{Erler:2019vhl,Okawa2006}.
It obeys the sliver gauge condition 
\begin{equation}
\mathcal{B}_{0}\Psi_{TV}=0,
\end{equation}
hence we want to use the associated stub operator $e^{-\lambda\mathcal{L}_{0}}$. 

\subsection{Action of $\mathbf{P}$\label{subsec:Action-of P}}

We now want to apply $\mathbf{P}$ onto $\Psi_{TV}$ since it is the
only cohomomorphism we have available in closed form. The linear term
is just $e^{-\lambda\mathcal{L}_{0}^{*}}\Psi_{TV}$, which can be
computed by expanding $\Psi_{TV}$ in formal eigenstates of $\mathcal{L}_{0}^{*}$
(see Appendix B). The resulting general formula is 
\begin{equation}
e^{-\lambda\mathcal{L}_{0}^{*}}\left(e^{-\alpha K}f\left(K,B,c\right)e^{-\alpha K}\right)=e^{-\left(\left(\alpha+\frac{1}{2}\right)e^{\lambda}-\frac{1}{2}\right)K}f\left(e^{\lambda}K,e^{\lambda}B,e^{-\lambda}c\right)e^{-\left(\left(\alpha+\frac{1}{2}\right)e^{\lambda}-\frac{1}{2}\right)K}
\end{equation}
from which 
\begin{align}
e^{-\lambda\mathcal{L}_{0}^{*}}\Psi_{TV}\, & =e^{-K\left(e^{\lambda}-\frac{1}{2}\right)}c\frac{KB}{1-e^{-e^{\lambda}K}}ce^{-K\left(e^{\lambda}-\frac{1}{2}\right)}\label{eq:stubbed tachyon vacuum}
\end{align}
follows.

To calculate the quadratic terms we first need to compute $h\Psi_{TV}$.
Here a major difference to the standard frame occurs because suppose
we have a solution in Siegel gauge and apply 
\begin{equation}
h=\frac{e^{-2\lambda L_{0}}-1}{L_{0}}b_{0},
\end{equation}
the result vanishes and hence the new solution is just given by the
linear term only. It is of course tempting to use our result for $h$
in sliver gauge and apply it to $\Psi_{TV}$ which results in 
\begin{equation}
h_{\mathcal{B}_{0}}\Psi_{TV}=\left(e^{-\lambda\mathcal{L}_{0}}\frac{\mathcal{B}_{0}^{*}}{\mathcal{L}_{0}^{*}}Q\frac{\mathcal{B}_{0}}{\mathcal{L}_{0}}e^{-\lambda\mathcal{L}_{0}^{*}}-\frac{\mathcal{B}_{0}^{*}}{\mathcal{L}_{0}^{*}}Q\frac{\mathcal{B}_{0}}{\mathcal{L}_{0}}\right)\Psi_{TV}.\label{eq:hPsi}
\end{equation}
However, as we have seen in the last section, the equations of motion
using $h_{\mathcal{B}_{0}}$exhibit singularities on certain states
including the zero-momentum tachyon $e^{-\frac{K}{2}}ce^{-\frac{K}{2}}$,
so it is not clear if we will end up with a well-defined solution.
Actually, $h_{\mathcal{B}_{0}}\Psi_{TV}$ is ill-defined as can be
shown as follows: We will focus on the first term here, the second
term is dealt with in section \ref{subsec:A-simpler-stubbed}.

The action of $\mathcal{B}_{0}$ can be determined by the formula
\begin{equation}
\mathcal{B}_{0}\left(e^{-\frac{K}{2}}\Psi e^{-\frac{K}{2}}\right)=e^{-\frac{K}{2}}B^{-}\Psi e^{-\frac{K}{2}}
\end{equation}
where $B^{-}=\frac{1}{2}\left(\mathcal{B}_{0}-\mathcal{B}_{0}^{*}\right)$
acts as a star algebra derivative and obeys 
\begin{equation}
B^{-}K=B,\,\,\,\,\,\,\,\,\,\,\,\,\,B^{-}B=0,\,\,\,\,\,\,\,\,\,\,\,\,\,\,B^{-}c=0,
\end{equation}
the result is 
\begin{equation}
\mathcal{B}_{0}e^{-\lambda\mathcal{L}_{0}^{*}}\Psi_{TV}=\left(1-e^{\lambda}\right)e^{-K\left(e^{\lambda}-\frac{1}{2}\right)}\left[\frac{KB}{1-e^{-e^{\lambda}K}},c\right]e^{-K\left(e^{\lambda}-\frac{1}{2}\right)}.\label{eq:not yet problematic}
\end{equation}
To apply $\frac{1}{\mathcal{L}_{0}}$ it is convenient to use the
Schwinger representation 
\begin{equation}
\frac{1}{\mathcal{L}_{0}}=\int_{0}^{\infty}dt\,\,e^{-t\mathcal{L}_{0}}
\end{equation}
and use an expansion in eigenstates of $\mathcal{L}_{0}$ (see Appendix
B). The calculation is then analogous to (\ref{eq:stubbed tachyon vacuum})
and results in 
\begin{equation}
\frac{\mathcal{B}_{0}}{\mathcal{L}_{0}}e^{-\lambda\mathcal{L}_{0}^{*}}\Psi_{TV}=\int_{0}^{\infty}dt\,\left(e^{-t}-e^{\lambda-t}\right)e^{-K\left(e^{\lambda-t}-e^{-t}+\frac{1}{2}\right)}\left[\frac{KB}{1-e^{-e^{\lambda-t}K}},c\right]e^{-K\left(e^{\lambda-t}-e^{-t}+\frac{1}{2}\right)}\label{eq:problematic term}
\end{equation}

This Schwinger integral is actually divergent: If we expand the integrand
for large $t$ we get 
\begin{equation}
\int_{0}^{\infty}dt\,\left(e^{-\lambda}-1\right)e^{-\frac{K}{2}}\left[B,c\right]e^{-\frac{K}{2}}
\end{equation}
which is an infinite integral over an expression independent of $t.$
The problem can also be seen from (\ref{eq:not yet problematic})
already: If we look at the $\mathcal{L}_{0}$-level expansion, we
see that the expression contains the term 
\begin{equation}
\left(e^{-\lambda}-1\right)e^{-\frac{K}{2}}\left[B,c\right]e^{-\frac{K}{2}}
\end{equation}
which has $\mathcal{L}_{0}$-eigenvalue zero and applying $\frac{1}{\mathcal{L}_{0}}$
is ill-defined. It is straightforward to show that the situation does
not improve by applying all the other operators in (\ref{eq:hPsi}),
especially since the divergence is not $Q$-closed: By isolating 
\begin{equation}
\left(h_{\mathcal{B}_{0}}\Psi_{TV}\right)_{div}=e^{-\lambda\mathcal{L}_{0}}\frac{\mathcal{B}_{0}^{*}}{\mathcal{L}_{0}^{*}}Q\frac{1}{0}\left(e^{-\lambda}-1\right)e^{-\frac{K}{2}}\left[B,c\right]e^{-\frac{K}{2}}
\end{equation}
and directly applying $Q$ we get 
\begin{equation}
Q\left(h_{\mathcal{B}_{0}}\Psi_{TV}\right)_{div}=e^{-\lambda\mathcal{L}_{0}}\frac{1}{0}\left(e^{-\lambda}-1\right)e^{-\frac{K}{2}}Q\left[B,c\right]e^{-\frac{K}{2}}=\frac{2}{0}\left(e^{-\lambda}-1\right)e^{-\frac{K}{2}}cKBce^{-\frac{K}{2}}\neq0.
\end{equation}

To get rid of the singularity we will instead use the vertices constructed
from $h_{\hat{\mathcal{B}}_{0}}$ in $\hat{\mathcal{B}_{0}}$-gauge
(\ref{eq:h in B_hat gauge}) given by 
\begin{equation}
h_{\hat{\mathcal{B}}_{0}}=\frac{e^{\left(e^{-\lambda}-1\right)\hat{\mathcal{\mathcal{L}}_{0}}}-1}{\hat{\mathcal{\mathcal{L}}_{0}}}\hat{\mathcal{\mathcal{B}}_{0}}=-\int_{0}^{1-e^{-\lambda}}dt\,e^{-t\hat{\mathcal{\mathcal{L}}_{0}}}\hat{\mathcal{\mathcal{B}}_{0}}.
\end{equation}
First we use the formula 
\begin{equation}
\hat{\mathcal{\mathcal{B}}_{0}}\Psi=B\Psi+\left(-\right)^{\text{gh}\left(\Psi\right)}\Psi B
\end{equation}
to write 
\begin{equation}
\hat{\mathcal{\mathcal{B}}_{0}}\Psi_{TV}=e^{-\frac{K}{2}}\left[\frac{KB}{1-e^{-K}},c\right]e^{-\frac{K}{2}}.
\end{equation}
Then we exponentiate the well-known relation 
\begin{equation}
\hat{\mathcal{\mathcal{L}}_{0}}\Psi=K\Psi+\Psi K\label{eq:L_hat relation}
\end{equation}
to get 
\begin{equation}
e^{-t\hat{\mathcal{\mathcal{L}}_{0}}}\Psi=e^{-tK}\Psi e^{-tK},\label{eq:exp(L_hat) relation}
\end{equation}
now the total action of $h$ becomes 
\begin{equation}
h_{\hat{\mathcal{B}}_{0}}\Psi_{TV}=-\int_{0}^{1-e^{-\lambda}}dt\,e^{-K\left(t+\frac{1}{2}\right)}\left[\frac{KB}{1-e^{-K}},c\right]e^{-K\left(t+\frac{1}{2}\right)}.
\end{equation}
As a second intermediate result we compute $ip\Psi_{TV}$ using (\ref{eq:h umformung})
and (\ref{eq:exp(L_hat) relation}): 
\begin{equation}
e^{-\lambda\mathcal{L}_{0}}e^{-\lambda\mathcal{L}_{0}^{*}}\Psi_{TV}=e^{-K\left(\frac{3}{2}-e^{-\lambda}\right)}c\frac{KB}{1-e^{-K}}ce^{-K\left(\frac{3}{2}-e^{-\lambda}\right)}.
\end{equation}
Combining the results and applying the projection operator $e^{-\lambda\mathcal{L}_{0}^{*}}$
once more yields the full quadratic part of the solution according
to (\ref{eq:P explicit}): 
\begin{align}
 & pm_{2}\left(\Psi_{TV},h_{\hat{\mathcal{B}}_{0}}\Psi_{TV}\right)+pm_{2}\left(h_{\hat{\mathcal{B}}_{0}}\Psi_{TV},ip\Psi_{TV}\right)\nonumber \\
=\, & \int_{0}^{1-e^{-\lambda}}dt\,e^{\lambda}(-e^{-K\left(e^{\lambda}-\frac{1}{2}\right)}c\frac{K^{2}B}{\left(1-e^{-e^{\lambda}K}\right)^{2}}e^{-K\left(e^{\lambda}t+e^{\lambda}\right)}ce^{-K\left(e^{\lambda}+e^{\lambda}t-\frac{1}{2}\right)}\\
 & +e^{-K\left(e^{\lambda}-\frac{1}{2}\right)}c\frac{KB}{1-e^{-e^{\lambda}K}}c\frac{K}{1-e^{-e^{\lambda}K}}e^{-K\left(2e^{\lambda}t+2e^{\lambda}-\frac{1}{2}\right)}\nonumber \\
 & -e^{-K\left(e^{\lambda}-\frac{1}{2}\right)}c\frac{KB}{1-e^{-e^{\lambda}K}}e^{-K\left(e^{\lambda}t+e^{\lambda}\right)}c\frac{K}{1-e^{-e^{\lambda}K}}e^{-K\left(e^{\lambda}+e^{\lambda}t-\frac{1}{2}\right)}\nonumber \\
 & +e^{-K\left(e^{\lambda}+e^{\lambda}t-\frac{1}{2}\right)}\frac{K}{1-e^{-e^{\lambda}K}}e^{-K\left(e^{\lambda}t+2e^{\lambda}-1\right)}c\frac{KB}{1-e^{-e^{\lambda}K}}e^{-K\left(2e^{\lambda}-\frac{3}{2}\right)}\nonumber \\
 & -e^{-K\left(e^{\lambda}+e^{\lambda}t-\frac{1}{2}\right)}\frac{K}{1-e^{-e^{\lambda}K}}ce^{-K\left(e^{\lambda}t+2e^{\lambda}-1\right)}\frac{KB}{1-e^{-e^{\lambda}K}}ce^{-K\left(2e^{\lambda}-\frac{3}{2}\right)}\nonumber \\
 & -e^{-K\left(e^{\lambda}+e^{\lambda}t-\frac{1}{2}\right)}c\frac{K^{2}B}{\left(1-e^{-e^{\lambda}K}\right)^{2}}e^{-K\left(e^{\lambda}t+2e^{\lambda}-1\right)}ce^{-K\left(2e^{\lambda}-\frac{3}{2}\right)}).\label{eq:complicated}
\end{align}
This expression is well-defined since (\ref{eq:h in B_hat gauge})
has no poles. An interesting point is that it fails to be twist symmetric,
so we conclude that the application of $\mathbf{P}$ in general breaks
twist symmetry.

\subsection{A simpler stubbed theory?\label{subsec:A-simpler-stubbed}}

Since the last result is quite complicated already at quadratic order
in $\Psi_{TV}$, one might ask if there exists a stubbed theory with
simpler solutions. For example one may try the following: Let us replace
$\mathcal{L}_{0}$ by $\mathcal{L}_{0}^{*}$ in every formula, i.
e. switch the operators $i$ and $p$ such that the new product becomes
\begin{equation}
M_{2}\left(\cdot,\cdot\right)=e^{-\lambda\mathcal{L}_{0}}\left(e^{-\lambda\mathcal{L}_{0}^{*}}\cdot,e^{-\lambda\mathcal{L}_{0}^{*}}\cdot\right)
\end{equation}
This choice is less natural than the original one since in the three
vertex only $\mathcal{L}_{0}^{*}$s would appear instead of the scaling
operator $\mathcal{L}_{0}$. However, the homotopy in sliver gauge
would become 
\begin{equation}
h'_{\mathcal{B}_{0}}=\left(e^{-\lambda\mathcal{L}_{0}^{*}}\frac{\mathcal{B}_{0}}{\mathcal{L}_{0}}Q\frac{\mathcal{B}_{0}^{*}}{\mathcal{L}_{0}^{*}}e^{-\lambda\mathcal{L}_{0}}-\frac{\mathcal{B}_{0}}{\mathcal{L}_{0}}Q\frac{\mathcal{B}_{0}^{*}}{\mathcal{L}_{0}^{*}}\right)P_{+}+\left(e^{-\lambda\mathcal{L}_{0}^{*}}\frac{\mathcal{B}_{0}^{*}}{\mathcal{L}_{0}^{*}}Q\frac{\mathcal{B}_{0}}{\mathcal{L}_{0}}e^{-\lambda\mathcal{L}_{0}}-\frac{\mathcal{B}_{0}^{*}}{\mathcal{L}_{0}^{*}}Q\frac{\mathcal{B}_{0}}{\mathcal{L}_{0}}\right)P_{-}
\end{equation}
so one would naively assume 
\begin{equation}
h'_{\mathcal{B}_{0}}\Psi_{TV}=0\label{eq:homotopy annihilates Psi TV}
\end{equation}
because of the gauge condition. However, there is an important caveat
here: $\Psi_{TV}$ contains the ghost number one state $-\frac{1}{2}e^{-\frac{K}{2}}cKBce^{-\frac{K}{2}}$
which is annihilated by $\mathcal{B}_{0}$ and $\mathcal{L}_{0}$,
so the result is ambiguous. To make sense of it it is useful to write
\begin{equation}
e^{-\frac{K}{2}}cKBce^{-\frac{K}{2}}=Q\left(e^{-\frac{K}{2}}Bce^{-\frac{K}{2}}\right)
\end{equation}
and use (\ref{eq:propagator rewriting}) such that we arrive at 
\begin{equation}
h'_{\mathcal{B}_{0}}\Psi_{TV}=\frac{1}{2}\left(1-e^{-\lambda\mathcal{L}_{0}^{*}}\right)\frac{\mathcal{B}_{0}^{*}}{\mathcal{L}_{0}^{*}}e^{-\frac{K}{2}}cKBce^{-\frac{K}{2}}.
\end{equation}
This result is also compatible with the Hodge-Kodaira relation (\ref{eq:Hodge Kodaira}).
Using the same techniques as above we compute 
\begin{equation}
h'_{\mathcal{B}_{0}}\Psi_{TV}=\frac{1}{2}\int_{0}^{\lambda}dt\,e^{t}e^{-K\left(e^{t}-\frac{1}{2}\right)}\left[KB,c\right]e^{-K\left(e^{t}-\frac{1}{2}\right)}
\end{equation}
and obtain to quadratic order in $\Psi_{TV}$ 
\begin{align}
\pi_{1}\mathbf{P}\frac{1}{1-\Psi_{TV}} & =e^{-\frac{K}{2}}c\frac{KB}{1-e^{-e^{-\lambda}K}}ce^{-\frac{K}{2}}\nonumber \\
 & \,\,\,+\frac{1}{2}\int_{0}^{\lambda}dt\,e^{t-\lambda}\big(e^{-\frac{K}{2}}c\frac{KB}{1-e^{-K}}e^{-e^{t-\lambda}K}\left\{ c,K\right\} e^{-K\left(e^{t-\lambda}-e^{-\lambda}+\frac{1}{2}\right)}\nonumber \\
 & \,\,\,-e^{-\frac{K}{2}}c\frac{KB}{1-e^{-K}}cKe^{-K\left(2e^{t-\lambda}-e^{-\lambda}+\frac{1}{2}\right)}\nonumber \\
 & \,\,\,+e^{-K\left(e^{t-\lambda}-e^{-\lambda}+\frac{1}{2}\right)}\left\{ c,K\right\} e^{-K\left(e^{t-\lambda}+e^{-\lambda}-e^{-2\lambda}\right)}\frac{KB}{1-e^{-K}}ce^{-K\left(\frac{1}{2}+e^{-\lambda}-e^{-2\lambda}\right)}\nonumber \\
 & \,\,\,-e^{-\left(2e^{t-\lambda}+\frac{1}{2}-e^{-2\lambda}\right)}Kc\frac{KB}{1-e^{-K}}ce^{-K\left(\frac{1}{2}+e^{-\lambda}-e^{-2\lambda}\right)}\big)+\cdots
\end{align}
This result is indeed a bit simpler than (\ref{eq:complicated}) but
also here no obvious simplification is visible.

\subsection{Action of $\mathbf{F}$\label{subsec:Action-of F}}

We can also ask what happens if we use the cyclic cohomomorphism (\ref{eq:cyclic intertwiner})
where the operator $X$ was defined in (\ref{eq:X}) and apply it
on $\Psi_{TV}$:
\begin{equation}
\pi_{1}\mathbf{F}\frac{1}{1-\Psi_{TV}}=\pi_{1}\mathcal{P}\text{exp}\left(\int_{0}^{\lambda}dt\,\left(\sum_{n=2}^{\infty}\pi_{1}\left(\mathbf{X^{*}M_{n}\left(\lambda\right)-M_{n}\left(\lambda\right)}\mathbf{X}\right)\pi_{n}\right)\right)\frac{1}{1-\Psi_{TV}}.
\end{equation}
At this stage the result is not very satisfying, but we can analyze
it for an infinitesimal $\lambda\ll1$: The sum collapses then to
the $n=2$ term and yields
\begin{equation}
\pi_{1}\mathbf{F}\frac{1}{1-\Psi_{TV}}=\Psi_{TV}+\lambda\left(X^{*}m_{2}\left(\Psi_{TV}^{\otimes2}\right)-m_{2}\left(X\Psi_{TV},\Psi_{TV}\right)-m_{2}\left(\Psi_{TV},X\Psi_{TV}\right)\right)+\mathcal{O}\left(\lambda^{2}\right).
\end{equation}
Let us again first try to use $h_{\mathcal{B}_{0}}$ for the vertices:
The first becomes explicitly 
\begin{equation}
X^{*}m_{2}\left(\Psi_{TV},\Psi_{TV}\right)=\frac{\mathcal{B}_{0}}{\mathcal{L}_{0}}Q\mathcal{B}_{0}^{*}m_{2}\left(\Psi_{TV},\Psi_{TV}\right).
\end{equation}
Using the equations motion for $\Psi_{TV}$ we can write this as 
\begin{equation}
-\frac{\mathcal{B}_{0}}{\mathcal{L}_{0}}Q\mathcal{B}_{0}^{*}Q\Psi_{TV}=-\frac{\mathcal{B}_{0}}{\mathcal{L}_{0}}Q\mathcal{L}_{0}^{*}\Psi_{TV}.
\end{equation}
The action of $\mathcal{L}_{0}^{*}$ is straightforward to calculate
using (\ref{eq:L_0 star relation}), the result is 
\begin{equation}
\mathcal{L}_{0}^{*}\Psi_{TV}=\left\{ K,e^{-\frac{K}{2}}c\frac{KB}{1-e^{-K}}ce^{-\frac{K}{2}}\right\} -e^{-\frac{K}{2}}c\frac{e^{-K}K^{2}B}{\left(1-e^{-K}\right)^{2}}ce^{-\frac{K}{2}}
\end{equation}
Applying $Q$ yields 
\begin{align}
Q\mathcal{L}_{0}^{*}\Psi_{TV}= & \,\left\{ K,e^{-\frac{K}{2}}cKc\frac{KB}{1-e^{-K}}ce^{-\frac{K}{2}}\right\} +\left\{ K,e^{-\frac{K}{2}}c\frac{KB}{1-e^{-K}}cKce^{-\frac{K}{2}}\right\} \nonumber \\
 & \,-\left\{ K,e^{-\frac{K}{2}}c\frac{K^{2}}{1-e^{-K}}ce^{-\frac{K}{2}}\right\} -e^{-\frac{K}{2}}cKc\frac{e^{-K}K^{2}B}{\left(1-e^{-K}\right)^{2}}ce^{-\frac{K}{2}}\nonumber \\
 & \,+e^{-\frac{K}{2}}c\frac{e^{-K}K^{3}}{\left(1-e^{-K}\right)^{2}}ce^{-\frac{K}{2}}-e^{-\frac{K}{2}}c\frac{e^{-K}K^{2}B}{\left(1-e^{-K}\right)^{2}}cKce^{-\frac{K}{2}}.
\end{align}
To check if we can apply $\frac{\mathcal{B}_{0}}{\mathcal{L}_{0}}$
we have to isolate the level zero contribution in the $\mathcal{L}_{0}$-expansion.
It is given by 
\begin{equation}
Q\mathcal{L}_{0}^{*}\Psi_{TV}\mid_{\mathcal{L}_{0}=0}=e^{-\frac{K}{2}}cKcKe^{-\frac{K}{2}}+e^{-\frac{K}{2}}KcKce^{-\frac{K}{2}}
\end{equation}
This state is non-vanishing and not annihilated by $\mathcal{B}_{0}$,
so the application of $\frac{\mathcal{B}_{0}}{\mathcal{L}_{0}}$ leads
again to an ill-defined result as anticipated.

Hence, to get something well-defined, we should use $h_{\hat{\mathcal{B}}_{0}}$
which leads to 
\begin{equation}
X\Psi_{TV}=\mathcal{B}_{0}\Psi_{TV}=0
\end{equation}
because of the gauge condition, so we arrive at 
\begin{equation}
\pi_{1}\mathbf{F}\frac{1}{1-\Psi_{TV}}=\Psi_{TV}+\lambda\mathcal{B}_{0}^{*}m_{2}\left(\Psi_{TV}^{\otimes2}\right)+\mathcal{O}\left(\lambda^{2}\right).
\end{equation}
This can be calculated explicitly using the same methods as in the
previous section: 
\begin{align}
\mathcal{B}_{0}^{*}m_{2}\left(\Psi_{TV}^{\otimes2}\right)= & \,\mathcal{B}_{0}^{*}\left(e^{-\frac{K}{2}}c\frac{KB}{1-e^{-K}}ce^{-K}c\frac{KB}{1-e^{-K}}ce^{-\frac{K}{2}}\right)\nonumber \\
= & \,e^{-\frac{K}{2}}\frac{KB}{1-e^{-K}}ce^{-K}c\frac{KB}{1-e^{-K}}ce^{-\frac{K}{2}}+e^{-\frac{K}{2}}c\frac{KB}{1-e^{-K}}ce^{-K}c\frac{KB}{1-e^{-K}}e^{-\frac{K}{2}}\nonumber \\
 & \,+e^{-\frac{K}{2}}c\frac{KB}{1-e^{-K}}e^{-K}c\frac{KB}{1-e^{-K}}ce^{-\frac{K}{2}}\nonumber \\
= & \,e^{-\frac{K}{2}}\left\{ \frac{KB}{1-e^{-K}}c\frac{K}{e^{K}-1},c\right\} e^{-\frac{K}{2}}+e^{-\frac{K}{2}}c\frac{K^{2}B}{e^{K}+e^{-K}-2}ce^{-\frac{K}{2}}\nonumber \\
 & -e^{-\frac{K}{2}}\left\{ \frac{KB}{e^{K}-1}c\frac{K}{1-e^{-K}},c\right\} e^{-\frac{K}{2}}.
\end{align}
In the end we get 
\begin{align}
\pi_{1}\mathbf{F}\frac{1}{1-\Psi_{TV}}=\, & e^{-\frac{K}{2}}c\frac{KB}{1-e^{-K}}ce^{-\frac{K}{2}}\nonumber \\
+\, & \lambda\,\big(e^{-\frac{K}{2}}\left\{ \frac{KB}{1-e^{-K}}c\frac{K}{e^{K}-1},c\right\} e^{-\frac{K}{2}}+e^{-\frac{K}{2}}c\frac{K^{2}B}{e^{K}+e^{-K}-2}ce^{-\frac{K}{2}}\nonumber \\
 & -e^{-\frac{K}{2}}\left\{ \frac{KB}{e^{K}-1}c\frac{K}{1-e^{-K}},c\right\} e^{-\frac{K}{2}}\big)\nonumber \\
+\, & \mathcal{O}\left(\lambda^{2}\right).
\end{align}
In contrast to our result for $\pi_{1}\mathbf{P}\frac{1}{1-\Psi_{TV}}$,
this expression is twist symmetric and from the discussion in section
\ref{subsec:The-symmetry-map} we also deduce that it is gauge equivalent
to (\ref{eq:complicated}) to first order in $\lambda.$ Indeed, comparing
the infinitesimal actions of $\mathbf{P}$ and $\mathbf{F}$ 
\begin{align}
\pi_{1}\mathbf{P}\frac{1}{1-\Psi_{TV}}\, & =\Psi_{TV}-\delta\lambda\mathcal{L}_{0}^{*}\Psi_{TV}-\delta\lambda\left(m_{2}\left(X^{*}\Psi_{TV},\Psi_{TV}\right)+m_{2}\left(\Psi_{TV},X^{*}\Psi_{TV}\right)\right)\\
\pi_{1}\mathbf{F}\frac{1}{1-\Psi_{TV}}\, & =\Psi_{TV}+\delta\lambda X^{*}m_{2}\left(\Psi_{TV},\Psi_{TV}\right)
\end{align}
one sees that with the definition $\Lambda=:X^{*}\Psi_{TV}$ the difference
can be written as 
\begin{equation}
\Delta\Psi_{TV}=\delta\lambda\left(Q\Lambda+m_{2}\left(\Lambda,\Psi_{TV}\right)+m_{2}\left(\Psi_{TV},\Lambda\right)+X^{*}\left(Q\Psi_{TV}+m_{2}\left(\Psi_{TV},\Psi_{TV}\right)\right)\right).\label{eq:combined trafo}
\end{equation}
Since $\Psi_{TV}$ obeys the original Witten's equation of motion,
$\Delta\Psi_{TV}$ is just a gauge transformation.

\section{Conclusion and outlook}

In this paper we addressed and answered some pending questions concerning
stubs in open string field theory and moreover generalized the whole
framework to more general gauges. We systematically constructed a
family of cohomomorphisms which map the Witten theory to the stubbed
theory, including the two maps $\mathbf{F^{-1}}$ and $\mathbf{P}$
already found in \cite{Schnabl:2023dbv}. To first order in $\lambda$,
the difference between two such maps always consists of a gauge transformation
and a part proportional to the equations of motion, according to the
conjecture of \cite{Schnabl:2023dbv} that the on-shell value of the
action is independent of the cohomomorphism. When moving to a different
coordinate frame we encountered some algebraic constraints on the
homotopy operator to get consistent and well-defined results. We have
shown that there is a solution for $h_{\mathcal{B}_{0}}$ under the
assumption that perturbation theory in the sliver frame is geometrically
consistent. While for on-shell amplitudes this operator $h_{\mathcal{B}_{0}}$
is well-dedfined, it exhibits singular behaviour on some off-shell
states in the Hilbert space. We concluded therefore that for the calculation
of classical solutions we should rather pick the $\hat{\mathcal{\mathcal{B}}_{0}}$-gauge
for the homotopy. This we could explicitly verify by computing the
tachyon vacuum in the stubbed theory up to second order. In a separate
calculation we also argued from a geometrical viewpoint why the inclusion
of higher vertices in the sliver frame is necessary. 

At this time it is not clear yet if those explicit expressions can
indeed lead us to more general properties of $A_{\infty}$-solutions
or even solutions of CSFT but it is certainly a possible direction
for further research. Especially the relation to \cite{Moeller2011},
where the closed string cohomology was identified in a purely open
string setup, is for sure worth to investigate. Another promising
path to follow is the connection to \cite{Erbin2023}, where stubs
in OSFT are represented through an auxiliary field. It would be interesting
if this formalism can be generalized to the sliver frame as well and
what the auxiliary field would look like for specific classical solutions.
In this context one could also make the role of twist symmetry more
precise and analyze under which conditions it can be preserved. Finally,
it would be worth to explicitly explore if the stubbed theory solves
issues with singularities, for instance concerning identity-based
solutions like $\Psi=c\left(1-K\right)$ \cite{Arroyo2010,Erler:2019vhl}.

\subsubsection*{Acknowledgements}

We thank Ted Erler, Jakub Vošmera, Martin Markl, Harold Erbin and
Atakan Firat for useful discussions. Our work has been funded by the
Grant Agency of Czech Republic under the grant EXPRO 20-25775X.

\appendix

\section{Tensor coalgebras}

The tensor coalgebra $TV$ associated to a (graded) vector space $V$
is defined as the Fock space 
\begin{equation}
V^{\otimes0}+V^{\otimes1}+V^{\otimes2}+...
\end{equation}
together with the comultiplication $\Delta:\,\,TV\rightarrow TV\otimes'TV$
given by 
\begin{equation}
\Delta\left(v_{1}\otimes...\otimes v_{n}\right)=\sum_{k=0}^{n}\left(v_{1}\otimes...\otimes v_{k}\right)\otimes'\left(v_{k+1}\otimes...\otimes v_{n}\right)
\end{equation}
on homogeneous elements and extended by linearity. Here the $v_{i}$
are elements of $V$ and $\otimes'$ denotes the tensor product arising
from a comultiplication, in contrast to the usual $\otimes$. We define
the projection operator $\pi_{n}:\,\,TV\rightarrow TV$ to project
any element on its $n$th tensor power component, 
\begin{equation}
\pi_{n}TV=V^{\otimes n}.
\end{equation}

A linear map $\mathbf{d}:\,\,TV\rightarrow TV$ is called a coderivation
if it satisfies the co-Leibniz rule: 
\begin{equation}
\Delta\mathbf{d}=\left(\mathbf{d}\otimes'\mathbf{1}+\mathbf{1}\otimes'\mathbf{d}\right)\Delta.\label{eq:co leibniz}
\end{equation}
Linear combinations of coderivations are again coderivations as well
as their graded commutator 
\begin{equation}
\left[\mathbf{d}_{1},\mathbf{d}_{2}\right]=\mathbf{d}_{1}\mathbf{d}_{2}-\left(-1\right)^{deg\left(\mathbf{d}_{1}\right)deg\left(\mathbf{d}_{2}\right)}\mathbf{d}_{2}\mathbf{d}_{1}.
\end{equation}
The product $\mathbf{d}_{1}\mathbf{d}_{2}$ is associative but in
general not a coderivation. For any $m$-linear map $d_{m}:\,\,V^{\otimes m}\rightarrow V$
one can construct an associated coderivation by the formula 
\begin{equation}
\mathbf{d_{m}}=\underset{n=m}{\overset{\infty}{\sum}}\underset{k=0}{\overset{n-m}{\sum}}1^{\otimes k}\otimes d_{m}\otimes1^{\otimes n-k-m}.\label{eq:coderivation}
\end{equation}
The co-Leibniz rule guarantees that any coderivation is a sum of terms
of this form for different $m$. Given two coderivations of this form,
we use the notation $\left(\mathbf{d}_{1}\mathbf{d}_{2}\right)$ to
denote the subset of terms where $d_{1m}$ acts on the output of $d_{2m'}$.
The individual $m$-products can be recovered from a general coderivation
as 
\begin{equation}
d_{m}=\pi_{1}\mathbf{d}\pi_{m}.
\end{equation}
If an odd coderivation $\mathbf{d}$ obeys 
\begin{equation}
\mathbf{d}^{2}=0
\end{equation}
then its components $d_{m}$ can be regarded as products of an $A_{\infty}$-algebra.

A linear map $\mathbf{f}$ is called a cohomomorphism if it fulfills
\begin{equation}
\Delta\mathbf{f}=\left(\mathbf{f}\otimes'\mathbf{f}\right)\Delta.
\end{equation}
Linear combinations and products of cohomomorphisms are again cohomomorphisms.
Given a family of $m$-products $f_{m}$ one can construct a unique
cohomomorphism via 
\begin{equation}
\mathbf{f}=\sum_{j=1}^{\infty}\sum_{k=1}^{\infty}\sum_{m_{1}+...+m_{j}=k}f_{m_{1}}\otimes...\otimes f_{m_{j}}.
\end{equation}
Again, the individual products can be recovered from $\mathbf{f}$
as 
\begin{equation}
f_{m}=\pi_{1}\mathbf{f}\pi_{m}.
\end{equation}

Of special importance are elements of $TV$ of the form 
\begin{equation}
1+v+v\otimes v+v\otimes v\otimes v+...=:\frac{1}{1-v}
\end{equation}
for some $v\in V.$ They fulfill the following useful properties:
\begin{equation}
\pi_{1}\mathbf{f}\frac{1}{1-v}=\sum_{m=1}^{\infty}f_{m}\left(v^{\otimes m}\right),
\end{equation}
\begin{equation}
\mathbf{f}\frac{1}{1-v}=\frac{1}{1-\pi_{1}\mathbf{f}\frac{1}{1-v}}\label{eq:cohomo relation}
\end{equation}
for any cohomomorphism $\mathbf{f}.$

A bilinear map $\Bra{\omega}$: $TV\times TV\rightarrow\mathbb{C}$
is called a symplectic form if it satisfies 
\begin{equation}
\Bra{\omega}\left(v_{1}\otimes v_{2}\right)=:\omega\left(v_{1},v_{2}\right)=-\left(-1\right)^{deg\left(v_{1}\right)deg\left(v_{2}\right)}\omega\left(v_{2},v_{1}\right).
\end{equation}
A multilinear product $m_{k}$ is called cyclic with respect to $\omega$
if it fulfills 
\begin{equation}
\omega\left(v_{1},m_{k}\left(v_{2},...,v_{k+1}\right)\right)=-\left(-1\right)^{deg\left(v_{1}\right)deg\left(m_{k}\right)}\omega\left(m_{k}\left(v_{1},...,v_{k}\right),v_{k+1}\right).
\end{equation}
A coderivation $\mathbf{d}$ is cyclic if all of its components $d_{m}=\pi_{1}\mathbf{d}\pi_{m}$
are cyclic or equivalently 
\begin{equation}
\Bra{\omega}\pi_{2}\mathbf{d}=0.
\end{equation}
Given two symplectic forms $\Bra{\omega}$, $\Bra{\omega'}$, a cohomomorphism
$\mathbf{f}$ is cyclic \footnote{See section 2.7 for a more careful treatment.}
if 
\begin{equation}
\Bra{\omega'}\pi_{2}\mathbf{f}=\Bra{\omega}\pi_{2}.
\end{equation}

\section{Eigenstates of $\mathcal{L}_{0}$ and $\mathcal{L}_{0}^{*}$}

The operator $\mathcal{L}_{0}$ obeys the familiar relation \cite{Erler:2019vhl}
\begin{equation}
\mathcal{L}_{0}\left(e^{-\frac{K}{2}}Xe^{-\frac{K}{2}}\right)=e^{-\frac{K}{2}}\frac{1}{2}\mathcal{L}^{-}Xe^{-\frac{K}{2}}\label{eq:L_0 action}
\end{equation}
where $\mathcal{L}^{-}=$$\mathcal{L}_{0}-\mathcal{L}_{0}^{*}$ and
$X$ is typically an element of the $KBc$-algebra. $\mathcal{L}^{-}$
acts as a derivation of the star algebra and fulfills 
\begin{equation}
\frac{1}{2}\mathcal{L}^{-}K=K,\,\,\,\,\,\,\,\,\,\,\,\,\,\frac{1}{2}\mathcal{L}^{-}B=B,\,\,\,\,\,\,\,\,\,\,\,\,\,\,\frac{1}{2}\mathcal{L}^{-}c=-c
\end{equation}
so we can deduce that states of the form 
\begin{equation}
e^{-\frac{K}{2}}K^{m}cK^{n}e^{-\frac{K}{2}}\,\,\,\,\,\,\text{or}\,\,\,\,\,e^{-\frac{K}{2}}K^{m}cK^{n}BcK^{r}e^{-\frac{K}{2}}
\end{equation}
are eigenstates of $\mathcal{L}_{0}$ with eigenvalue $m+n-1$ or
$m+n+r-1$, respectively.

To find the eigenstates of $\mathcal{L}_{0}^{*}$ we notice that equation
(\ref{eq:L_0 action}) can be alternatively written as 
\begin{equation}
\mathcal{L}_{0}X=\frac{1}{2}\mathcal{L}^{-}X+\frac{1}{2}\left(KX+XK\right),
\end{equation}
since the anticommutator with $K$ cancels the terms coming from the
action of $\frac{1}{2}\mathcal{L}^{-}$ on the security strips $e^{-\frac{K}{2}}$.
Now writing 
\begin{equation}
-\mathcal{L}_{0}^{*}X=\frac{1}{2}\mathcal{L}^{-}X-\frac{1}{2}\left(KX+XK\right)\label{eq:L_0 star relation}
\end{equation}
we see that the sign of $K$ in the security strips has to change.
Indeed, by a straightforward calculation one can show that 
\begin{equation}
e^{\frac{K}{2}}K^{m}cK^{n}e^{\frac{K}{2}}\,\,\,\,\,\,\text{or}\,\,\,\,\,e^{\frac{K}{2}}K^{m}cK^{n}BcK^{r}e^{\frac{K}{2}}
\end{equation}
are formal eigenstates of $\mathcal{L}_{0}^{*}$ with eigenvalue $-m-n+1$
or $-m-n-r+1$, respectively. They are in fact ill-defined due to
the appearance of inverse wedge states. However, we can still use
them as a formal device to determine the action of $\mathcal{L}_{0}^{*}$
on the string fields in question and get a well-defined final result.
So for example the tachyon vacuum can be expanded as 
\begin{equation}
\Psi_{TV}=\sum_{m,n,r=0}^{\infty}\frac{\left(-1\right)^{m+n+r}B_{n}}{m!n!r!}\left(e^{\frac{K}{2}}K^{m}cBK^{n}cK^{r}e^{\frac{K}{2}}\right).
\end{equation}
We can derive the two useful formulas 
\begin{equation}
e^{-\lambda\mathcal{L}_{0}}\left(e^{-\alpha K}f\left(K,B,c\right)e^{-\alpha K}\right)=e^{-\left(\left(\alpha-\frac{1}{2}\right)e^{-\lambda}+\frac{1}{2}\right)K}f\left(e^{-\lambda}K,e^{-\lambda}B,e^{\lambda}c\right)e^{-\left(\left(\alpha-\frac{1}{2}\right)e^{-\lambda}+\frac{1}{2}\right)K}
\end{equation}
\begin{equation}
e^{-\lambda\mathcal{L}_{0}^{*}}\left(e^{-\alpha K}f\left(K,B,c\right)e^{-\alpha K}\right)=e^{-\left(\left(\alpha+\frac{1}{2}\right)e^{\lambda}-\frac{1}{2}\right)K}f\left(e^{\lambda}K,e^{\lambda}B,e^{-\lambda}c\right)e^{-\left(\left(\alpha+\frac{1}{2}\right)e^{\lambda}-\frac{1}{2}\right)K}
\end{equation}

As a crosscheck we can verify the algebra 
\begin{equation}
\left[\mathcal{L}_{0},\mathcal{L}_{0}^{*}\right]=\mathcal{L}_{0}+\mathcal{L}_{0}^{*}
\end{equation}
by acting on an arbitrary function of $K$ and using the eigenstates:
\begin{equation}
\mathcal{L}_{0}^{*}F\left(K\right)=\mathcal{L}_{0}^{*}e^{\frac{K}{2}}\sum_{n}f_{n}K^{n}e^{\frac{K}{2}}=-\sum_{n}nf_{n}K^{n}e^{K}
\end{equation}
if $F\left(K\right)$ is expanded as $F\left(K\right)=\sum_{n}f_{n}K^{n}e^{K}$.
On the other hand, 
\begin{equation}
\mathcal{L}_{0}F\left(K\right)=\frac{1}{2}\mathcal{L}^{-}F\left(K\right)+KF\left(K\right)=K\left(F\left(K\right)+F'\left(K\right)\right)=2\sum_{n}f_{n}K^{n+1}e^{K}+\sum_{n}nf_{n}K^{n}e^{K}.
\end{equation}
Proceeding in the same way yields 
\begin{equation}
\mathcal{L}_{0}\mathcal{L}_{0}^{*}F\left(K\right)=-\mathcal{L}_{0}\sum_{n}nf_{n}K^{n}e^{K}=-\sum_{n}n^{2}f_{n}K^{n}e^{K}-2\sum_{n}nf_{n}K^{n+1}e^{K}
\end{equation}
\begin{equation}
\mathcal{L}_{0}^{*}\mathcal{L}_{0}F\left(K\right)=-2\sum_{n}\left(n+1\right)f_{n}K^{n+1}e^{K}-\sum_{n}n^{2}f_{n}K^{n}e^{K}.
\end{equation}
Now one can straightforwardly see that 
\begin{equation}
\left[\mathcal{L}_{0},\mathcal{L}_{0}^{*}\right]F\left(K\right)=2\sum_{n}f_{n}K^{n+1}e^{K}=\mathcal{L}_{0}F\left(K\right)+\mathcal{L}_{0}^{*}F\left(K\right)
\end{equation}
as expected. An alternative way to show that is to use the fact that
on functions of $K$, $\mathcal{L}_{0}$ and $\mathcal{L}_{0}^{*}$
can be represented as $K\frac{d}{dK}+K$ and $-K\frac{d}{dK}+K$ respectively,
and use 
\begin{equation}
\left[K\frac{d}{dK}+K,-K\frac{d}{dK}+K\right]=2K.
\end{equation}

\bibliographystyle{plain}
\bibliography{More_on_stubs}

\end{document}